\documentclass[useAMS,usenatbib]{mn2e}
\usepackage{epsfig,amsmath,amssymb}
\usepackage{natbib}
\usepackage{bm}
\usepackage{multirow}

\usepackage{graphicx,rotate,url,mathptmx,lscape,times,color}

\def \alii {Al\,{\sc ii}~}
\def \mgii {Mg\,{\sc ii}~}
\def \mgiins {Mg\,{\sc ii}}
\def \feii {Fe\,{\sc ii}~}

\def \civ {C\,{\sc iv}~}
\def \civns {C\,{\sc iv}}

\def \cv {C\,{\sc v}~}
\def \hi {H\,{\sc i}~}
\def \heii {He\,{\sc ii}~}
\def \heiins {He\,{\sc ii}}
\def \hins {H\,{\sc i}}
\def \nv {N\,{\sc v}~}
\def \nvns {N\,{\sc v}}
\def \nev {Ne\,{\sc v}~}
\def \oiv {O\,{\sc iv}~}
\def \ov {O\,{\sc v}~}
\def \ovi {O\,{\sc vi}~}
\def \ovins {O\,{\sc vi}}
\def \siiv {Si\,{\sc iv}~}

\def \ll {$ \lambda\lambda$}
\def \l {$ \lambda$}

\def \ebv {\ensuremath{E(B-V)}}

\def \apj {ApJ}

\def \nat {Nat}
\def \anips {Advanced Neural Inf. Processing Syst.}

\def \kms{\ensuremath{{\rm \,km\,s}^{-1}}}
\def \cm2{\ensuremath{{\rm \,cm}^{-2}}}
\usepackage{verbatim}

\title[Line-driven radiative outflows in luminous quasars]{Line-driven radiative outflows in luminous quasars}
\author[R. A. A. Bowler, P. C. Hewett, J. T. Allen and G. J. Ferland]
{Rebecca A. A. Bowler$^{1,2}$\thanks{E-mail: raab@roe.ac.uk}, 
Paul C. Hewett$^{1}$,
James T. Allen$^{1,3}$ and
Gary J. Ferland$^{4,5}$\\
1. Institute of Astronomy, University of Cambridge, Cambridge CB3 0HA \\
2. SUPA\thanks{Scottish Universities Physics Alliance}, Institute for Astronomy, University of Edinburgh, Royal Observatory,
Edinburgh EH9 3HJ \\
3. Sydney Institute for Astronomy, School of Physics, A28, University
of Sydney, NSW 2006, Australia \\
4. Department of Physics and Astronomy, University of Kentucky, 
Lexington, KY 40506, USA \\
5. Centre for Theoretical Atomic, Molecular and Optical Physics,
School of Mathematics and Physics,
Queens University Belfast, Belfast BT7 1NN, UK\\}

\begin{document}

%\date{Accepted 2010 December 15. Received 2010 December 14; in original form 2010 October 11}

%\pagerange{\pageref{firstpage}--\pageref{lastpage}} \pubyear{2012}

\maketitle

\label{firstpage}

\begin{abstract}
An analysis of $\simeq$19\,500 narrow ($\la$200\kms)
\civ$\lambda\lambda$1548.2,1550.8 absorbers in $\simeq$34\,000 Sloan Digital Sky
Survey quasar spectra is presented. The statistics of the number of
absorbers as a function of outflow-velocity shows that in
approximately two-thirds of outflows, with multiple \civ absorbers
present, absorbers are line-locked at the 500\kms velocity separation
of the \civ absorber doublet; appearing as `triplets' in the quasar
spectra.  Line-locking is an observational signature of radiative line
driving in outflowing material, where the successive shielding of
`clouds' of material in the outflow locks the clouds together in
outflow velocity. Line-locked absorbers are seen in both broad
absorption line quasars (BALs) and non-BAL quasars with comparable
frequencies and with velocities out to at least 20\,000\kms.  There
are no detectable differences in the absorber properties and the dust
content of single \civ doublets and line-locked \civ doublets.  The
gas associated with both single and line-locked \civ absorption
systems includes material with a wide range of
ionization potential (14-138\,eV). Both single and line-locked \civ
absorber systems show strong systematic trends in their ionization as
a function of outflow velocity, with ionization decreasing rapidly
with increasing outflow velocity.  Initial simulations, employing {\sc
Cloudy}, demonstrate that a rich spectrum of line-locked signals at
various velocities may be expected due to significant opacities from
resonance lines of Li-, He- and H-like ions of O, C and N, along with
contributions from \heii and \hi resonance lines. The simulations
confirm that line driving can be the dominant acceleration mechanism
for clouds with $N$(\hins)$\simeq$10$^{19}$\cm2.
\end{abstract}


\section{Introduction}\label{intro}

Evidence for highly ionized outflowing material is seen in the
rest-frame ultra-violet spectral energy distributions (SEDs) of many
quasars. Outflows are most prominent in broad absorption line quasars
(BALQSOs), where extensive line absorption by matter at outflow
velocities up to 30\,000\,\kms\ is seen as a broad trough blueward of
the rest-frame emission from the quasar
\citep{1991ApJ...373...23W}. BALQSOs themselves are observed to make
up some 20\,per cent of the population
\citep[e.g.][]{2003AJ....125.1784H} and the intrinsic fraction of
BALQSOs is even higher at redshifts $z\ga 3$
\citep{2011MNRAS.410..860A}. The ubiquitous presence of outflows with
velocities up to 12\,000\,\kms, visible via their absorption in the
\civns\,\ll1548.19,1550.77 doublet\footnote{Vacuum
  wavelengths are used throughout the paper}
\citep{2008MNRAS.386.2055N, 2008MNRAS.388..227W} suggests the estimate
that 60\,per cent of quasars possess outflows in the review by
\citet{2008ApJ...672..102G} may be conservative.

While the evidence for the presence of outflows associated with the
majority of quasars is established, the origin, dynamics and
composition of such outflows are poorly constrained. Models based on
radiation driven flows \citep{1985ApJ...294...96S,
  1994ApJ...432...62A, 1995ApJ...451..498M, 2000ApJ...543..686P,
  2001MNRAS.326..916C}, magnetohydrodynamic driven winds
\citep{1982MNRAS.199..883B, 1992ApJ...385..460E, 1994ApJ...434..446K,
  1997ApJ...479..200B} and thermal winds \citep{2007ASPC..373..267P}
all remain viable and it is possible that more than one mechanism is
important.

The quantity of material expelled from the central region of the
quasar, and the energy it carries, has important implications for
determining the lifetime of the quasar, the composition of the
interstellar medium (ISM) and galactic feedback mechanisms.
\citet{2010ApJ...722..642O} provide a recent review of the key
physical quantities involved.

Information about the nature of outflows, and the importance of
radiative driving in their generation, can be gained from an analysis
of the observed velocity structures present in the outflowing
material. Radiation-driving, and particularly the importance of
acceleration via the interaction of photons with bound electrons in
the outflowing material, is perhaps most familiar as the mechanism for
producing winds from hot stars. \citet{1975ApJ...195..157C} presented
the theory for stars at a time when spectroscopic observations of
quasars were just beginning to provide observational constraints on
the properties of gaseous material associated with the objects
\citep{1975ApJ...202..287B}.

Compared to the situation with hot stars, the geometry of outflows in
quasars is likely more complex and there are key differences in the
physical properties of the outflowing material, notably the much lower
particle densities. The theory behind radiatively-driven winds for
quasars was developed by Arav and coworkers in the 1990s
\citep{1994ApJ...427..700A, 1994ApJ...432...62A}. The importance of
radiation-driving for outflows in BALQSOs was stressed
\citep{1995Nature.376..576A} and \citet{1996ApJ...465..617A} proposed
the detection of the `ghost of Ly$\alpha$' as unambiguous evidence for
radiation-driving as the dominant source of acceleration in BALQSO
winds. The phenomenon is postulated to arise in outflows where
absorption of Ly$\alpha$ photons by \nvns-ions at a velocity of
$\simeq$5900\kms (where, in the frame of the outflowing \nvns-ions,
strong resonant absorption occurs) results in the rapid acceleration
of material to greater velocities, reducing the amount of material
present. The overall absorption opacity at $\simeq$5900\kms is thereby
reduced and an excess of emission is visible (e.g. see figure 1 of~\citealp{1996ApJ...465..617A}). For a tracer-ion in the
flow, such as \civns, the manifestation of the phenomenon is a reduction
in the absorption at $\simeq$5900\kms, leading to an apparent local
excess of continuum emission; hence the `ghost of Ly$\alpha$' name.

The in-depth consideration of the phenomenon by
\citet{1996ApJ...465..617A} was motivated in part by possible differences in
the spectra of non-BALQSOs and BALQSOs presented by
\citet{1991ApJ...373...23W} and the subsequent investigation of
\citet{1993ApJS...88..357K}. As discussed by
\citet{1996ApJ...465..617A}, very specific conditions are necessary for
the `ghost of Ly$\alpha$' to be detected. An outflow, including
significant \nv absorption, extending to $\sim$10\,000\kms, must be
present in a quasar that possesses both strong and relatively narrow
intrinsic Ly$\alpha$ emission. The ghost signature is predicted to be
most obvious for quasars where the far-ultraviolet spectrum of the
quasar is weak, suppressing any contribution to the acceleration of the
outflow from additional absorption due to resonance lines with
wavlengths $<$1000\AA.

Observational evidence for the presence of the ghost of Ly$\alpha$
has proved elusive, even when a systematic approach, using the power
of the very large sample of homogeneous, moderate
resolution and moderate signal-to-noise ratio spectra of quasars
\citep{2010AJ....139.2360S} from the Sloan Digital Sky Survey (SDSS)
are employed \citep{2010MNRAS.406..2094C}.

In this paper, we undertake a complementary search, focused on the
presence of `line-locked' absorption features in the spectra of
individual quasars with extended narrow line absorption. Such
line-locked absorption is also accepted as evidence that radiative
line-driving plays an important role in quasar outflow
dynamics~\citep{2005MNRAS.360.1455B, 1987ApJ...317..450F}.

The principle behind line-locking in radiatively-driven outflows is
that, under certain circumstances, material closer to the origin of
the outflow can shield material further out, resulting in a
synchronising of the outflow velocities. Specifically, if a distant
`cloud' is being accelerated outwards by line-driving arising from
resonant absorption of photons in a particular ionic species, the
reduction in the flux received by this cloud due to shielding by an
inner cloud, moving with a relative velocity equivalent to the
separation of two prominent absorption lines, could cause the clouds
to become locked together in velocity~\citep{1926MNRAS..86..459M,
1973ApJ...179..705S, 1989ApJ...342..100B}.  Simulations of the
structure, dynamics and origin of quasar outflows also indicate that,
for radiatively driven flows, line-locking could be
prevalent~\citep{2003MNRAS.344..233C}. Quasar outflows are almost
certainly quite complex and simulations
\citep[e.g.][]{2008ApJ...676..101P, 2009ApJ...693.1929K,
2014ApJ...780...51P} predict the presence of significant structure on
a variety of scales, involving fragmentation and clumping, some of
which may be highly non-spherical (e.g. filamentary or
pancake-like). Our use of the 'cloud'-terminology above, and later in
the paper, is as a shorthand for some form of modulation in density
within the flow, along the line-of-sight to the illuminating source.

The potential significance of observing a clear signature of
line-locking has been recognised \citep{1975ApJ...202..287B} for much
of the period since the discovery of quasars. Evidence for the
presence of line-locking signatures have, however, been largely
confined to the identification of `picket fence'-type absorption
configurations in a relatively small number of individual
quasars~\citep[e.g.][]{1975ApJ...202..296W, 1976ApJ...207....1C,
1978ApJ...223..758A, 1987ApJ...317..450F, 1996AJ....111.1456B,
1997ApJS..112....1T, 2000ApJ...528..617S, 2004A&A...418..857F,
2005MNRAS.360.1455B, 2010MNRAS.409..269S, 2011MNRAS.410.1957H}.

The availability of the SDSS spectroscopic quasar sample 
\citep{2010AJ....139.2360S} now allows quantitative statistical 
investigation of outflow absorption signatures to be undertaken.
We focus on the properties of absorption due to 
triply-ionized carbon, specifically the doublet transition 
\civns\,\ll1548.19,1550.77, which is the most readily observable
high-ionization absorption feature in the SDSS spectra that appears
longward of the effects of the Ly$\alpha$-forest (below $\simeq$1200\AA).

The nature of \civ absorption in BALQSOs and non-BALQSOs is
dramatically different and we analyse samples of BALQSOs and
non-BALQSOs separately. Initially we began the investigation using an
auto-correlation calculation, designed to be sensitive to the presence
of many, individually difficult to detect, faint absorption
features. More specifically, we looked for correlations between
`absorber pixels' as a function of velocity separation, within each
quasar spectrum, without the need to pre-identify individual, discrete
absorbers. The auto-correlation method was devised primarily for
BALQSOs, which represent a subset of the quasar population where
outflows of material are very definitely present, yet the complex
absorption profiles of the BAL-troughs mean that searches for discrete
absorbers might not represent the optimum approach. In practice, such
a reservation turned out to be unfounded and, as a result, we employed
the familiar approach of identifying discrete absorption features and
then investigating the incidence of absorber `pairs' as a function of
velocity separation. The discrete-absorber method turns out to be
highly effective for the investigation of the properties of
individual, relatively narrow, \civns-absorbers present in both the
non-BALQSO and BALQSO quasar samples.

The paper is structured as follows. First, we present the SDSS quasar
sample in Section~\ref{sec:quasarsample}.  The absorption line
identification scheme is described in
Section~\ref{sec:discabs}. Statistics of the population of
\civns-absorbers as a function of outflow velocity are presented in
Section~\ref{sec:ll_stats}.  In Section 5, we create and analyse
composite quasar spectra as a function of outflow velocity, allowing
an investigation into the physical properties of the absorbers and the
host quasars. Line-locking is shown to be almost ubiquitous among
narrow \civns-absorbers in quasar outflows with multiple absorbers., A
discussion of the interpretation of the results, including exploratory
simulations using {\sc Cloudy}, is presented in
Section~\ref{sec:discuss}, before the main conclusions of the paper
are summarised in Section~\ref{sec:conclude}.

\section{The Quasar Sample} \label{sec:quasarsample}

The base quasar sample consisted of 42\,273 objects with redshifts
$1.54 \le z \le 3.0$ from the DR7 quasar catalogue of
\citet{2010AJ....139.2360S}, where the lower redshift limit was
imposed in order for the blue-wing of the \civ emission line to lie
within the SDSS spectra. Above redshift $z=3.0$ the number of quasars
in the SDSS catalogue drops rapidly, and the majority are fainter than
their lower redshift counterparts resulting in a significant reduction
in the signal-to-noise ratio (S/N) of the spectra. The quasars possess
absolute magnitudes in the range -23.0$\le$$M_i$$\le$-29.5 
\citep{2011ApJS..194...45S} but the sample is predominantly (79\,per
cent) made up of objects with -25.0$\le$$M_i$$\le$-27.0.

The spectra were all processed through the sky-residual subtraction
scheme of \citet{2005MNRAS.358.1083W}, which results in significantly
improved S/N at observed wavelengths \l$>$7200\,\AA,
contributing to the estimation of more accurate quasar
redshifts\footnote{http://www.sdss.org/dr7/products/value\_added/index.html/\#quasars}
from \citet{2010MNRAS.405.2302H}, which are employed throughout. For
3581 quasars in the sample, the SDSS DR7 release includes multiple
spectra of high quality. For these quasars, the spectra were co-added
using inverse variance weighting to produce a combined spectrum with
improved S/N.  All spectra were also corrected for the effects of
Galactic extinction using the SDSS \ebv\ measurement, taken from the
dust maps of \citet*{1998ApJ...500..525S}, and the Milky Way
extinction curve of \citet*{1989ApJ...345..245C}.

The sample was culled to remove objects with spectra of low S/N in
which the majority of moderate equivalent width \civ absorption
features are not readily detectable. Specifically, spectra were
retained in the sample if the average S/N per pixel in the continuum
redward of the \civ emission, at 1600-1650\,\AA, exceeded S/N=5.0. To
eliminate objects possessing spectra with extended wavelength regions
missing, spectra with fewer than 3500 pixels (defined by the SDSS-flag
{\sc NGOOD}) were also excluded. Application of the two criterion
resulted in the removal of 8477 spectra, leaving a sample of 33\,796
quasars.

Samples of non-BALQSOs and BALQSOs were defined using the
classifications of \cite{2011MNRAS.410..860A} for quasars present in
the SDSS DR6 release. The small fraction of quasars present only in
DR7 were classified via visual inspection\footnote{None of the results
presented are sensitive to the exact definition of the BALQSO and
non-BALQSO samples.}. The resulting sub-samples of non-BALQSOs and 
BALQSOs contained 31\,142 and 2654 quasars respectively.

The SDSS spectra are supplied on a logarithmic wavelength scale, with
pixels of $69$~\kms\, width. When defining reference
wavelengths, or generating composite spectra in the quasar rest-frame,
nearest-pixel values were employed and no interpolation or rebinning
was performed at any stage.

\section{The Absorber Sample}\label{sec:discabs}

\subsection{Absorption line detection}

Discrete absorption features were identified using the following
scheme.  A `continuum' is defined for each quasar via the application
of a simple 41--pixel median filter.  The `difference' spectrum, to be
searched for absorption features, is then obtained by subtracting the
continuum from the original quasar spectrum.  The absorption line
search uses a matched-filter technique
\citep[e.g.][]{1985MNRAS.213..971H} with template Gaussian profiles of
full width at half maximum (FWHM) = 160\kms (the resolution of the
SDSS spectra), 200\kms and 240\kms. Templates with lines centred at
half-pixel (i.e. 34.5\kms) intervals are employed, allowing the
absorber centroids to be located to better than $\pm$34.5\kms. At each
half-pixel, the template giving the minimum $\chi^2$ value is
determined and candidate absorbers are selected by applying a
detection-threshold of S/N$\ge$3, where the noise is calculated using
the array of variance values included with each SDSS spectrum. The low
value of the S/N-threshold is chosen deliberately to ensure the
inclusion of as many real features as possible, albeit at the expense
of the presence of a population of spurious `detections'.

The combination of the finite resolution of the quasar spectra and the
intrinsic velocity widths of absorption systems precludes the
detection of pairs of absorbers with small velocity separations. Two
unresolved absorbers must be at least $\simeq$5-pixels
($\simeq$345\kms) apart to be detected as distinct
features\footnote{A more sophisticated, multi-component scheme to
identify closer absorbers is in principle possible, but neither the
scientific goal or the quality of the spectra merit such an approach.}
and the `exclusion' interval becomes somewhat larger for features that
are resolved, i.e. possess intrinsic velocity widths $\ga$100\kms. A
\civ doublet was defined from the presence of two absorption features
separated by 7.24$\pm$1.00 pixels.  The \civ doublet velocity
separation of 500\kms (i.e. 7.24 pixels in the SDSS spectra) is such
that \civ absorbers with intrinsic velocity widths of up to
$\simeq$200\kms are detectable, hence the maximum template velocity
width of 240\kms used for absorber detection. Real \civ doublets with
larger observed velocity widths no longer appear as two separate,
individual, absorbers and are thus not included in the doublet catalogue.  

\begin{figure}
\includegraphics[width=84mm]{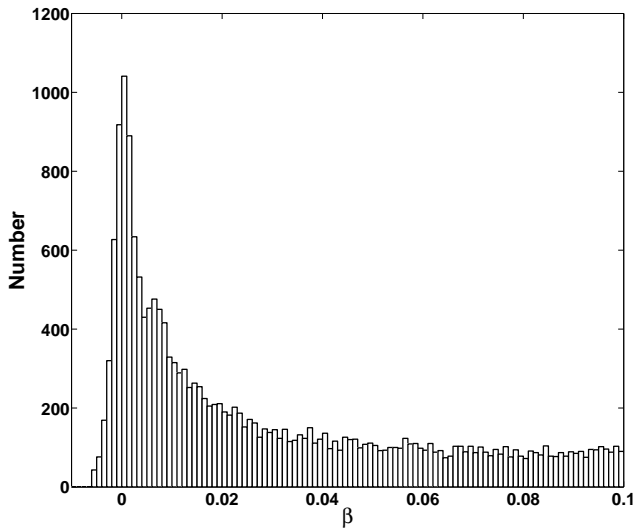}
 \caption{The distribution of the full sample of 18\,908 \civ
absorbers as a function of outflow velocity, $\beta = v/c$, identified
in the sample of 31\,142 non-BALQSOs. The majority of \civ absorbers
with $\beta > 0.05$ are expected to arise in intervening galaxy
haloes, unrelated to the quasars. The excess of apparently outflowing
absorbers is present out to at least $\beta=0.04$. An excess of \civ
absorbers with a relatively low velocity-dispersion
($\sigma_{observed} \simeq 500\kms$---including remaining quasar
redshift errors) centred on the quasar systemic velocity is also
evident from employing the improved quasar redshifts of
\citet{2010MNRAS.405.2302H}.
}
 \label{f:betadist}
\end{figure}

The rest-frame equivalent width (EW$_{1548}$)-limit reached depends on
the S/N of the individual quasar spectra. The absorber sample is
essentially complete for the EW$_{1548}$-interval $0.4 < {\rm
EW}_{1548} < 1.0$\,\AA. At lower EW$_{1548}$, the sample becomes
increasingly incomplete; the number of absorbers detected reaches a
maximum at EW$_{1548}$=0.3\,\AA, below which the number of systems
detected declines precipitately towards the minimum EW included in the
non-BALQSO sample of EW$_{1548}$=0.1\,\AA. Above EW$_{1548}$=1.0\,\AA
\ the sample also rapidly becomes incomplete because of the maximum
velocity width (FWHM = 240\kms) of the detection templates, imposed to
ensure that the components of the \civ doublets remain resolved.

For the BALQSO sample the catalogue of narrow absorbers is less
complete because of the presence of significant absorption, of
substantial optical depth, over much larger velocity intervals. Less
than 1\, percent of the \civ doublets were excluded from the
non-BALQSO sample by applying the EW$_{1548}$$\ge$0.1\,\AA \
condition, and, given the difficulty of defining a meaningful
EW-estimate for many of the narrow absorbers lying within broader
absorption features, no EW$_{1548}$-limit was applied to the absorber
catalogue from the BALQSO sample.

Investigation of the number of line-locked \civ systems within the
absorber catalogue is differential in nature and determination of the
exact completeness of the \civ absorber sample as a function of
EW$_{1548}$ does not affect the results presented. The key factor is
that the relative probabilities of detecting \civ doublets and \civ
triplets (i.e. potentially line-locked systems) are well
determined. None of the results presented in the paper change, other
than the exact number of absorbers involved, if the absorber sample is
restricted to minimum EW$_{1548}$ thresholds of up to
0.4\,\AA. Results from the full sample, to a minimum
EW$_{1548}$=0.1\,\AA, are presented to maximise the number of \civ
absorbers involved and hence the S/N of the results for both doublet
and triplet absorbers.

The absorber sample has properties very similar to other \civ absorber
samples derived recently from SDSS spectra
\citep[e.g.][]{2013ApJ...763...37C}, although the scheme for continuum
determination and feature finding are somewhat different. The apparent
optical depth method \citep{1991ApJ...379..245S} leads to an estimate
of the minimum column densities reached of $\rm{log}(N_{C\,{\sc
IV}})\simeq 14.0$.

\begin{figure}
\includegraphics[width=84mm]{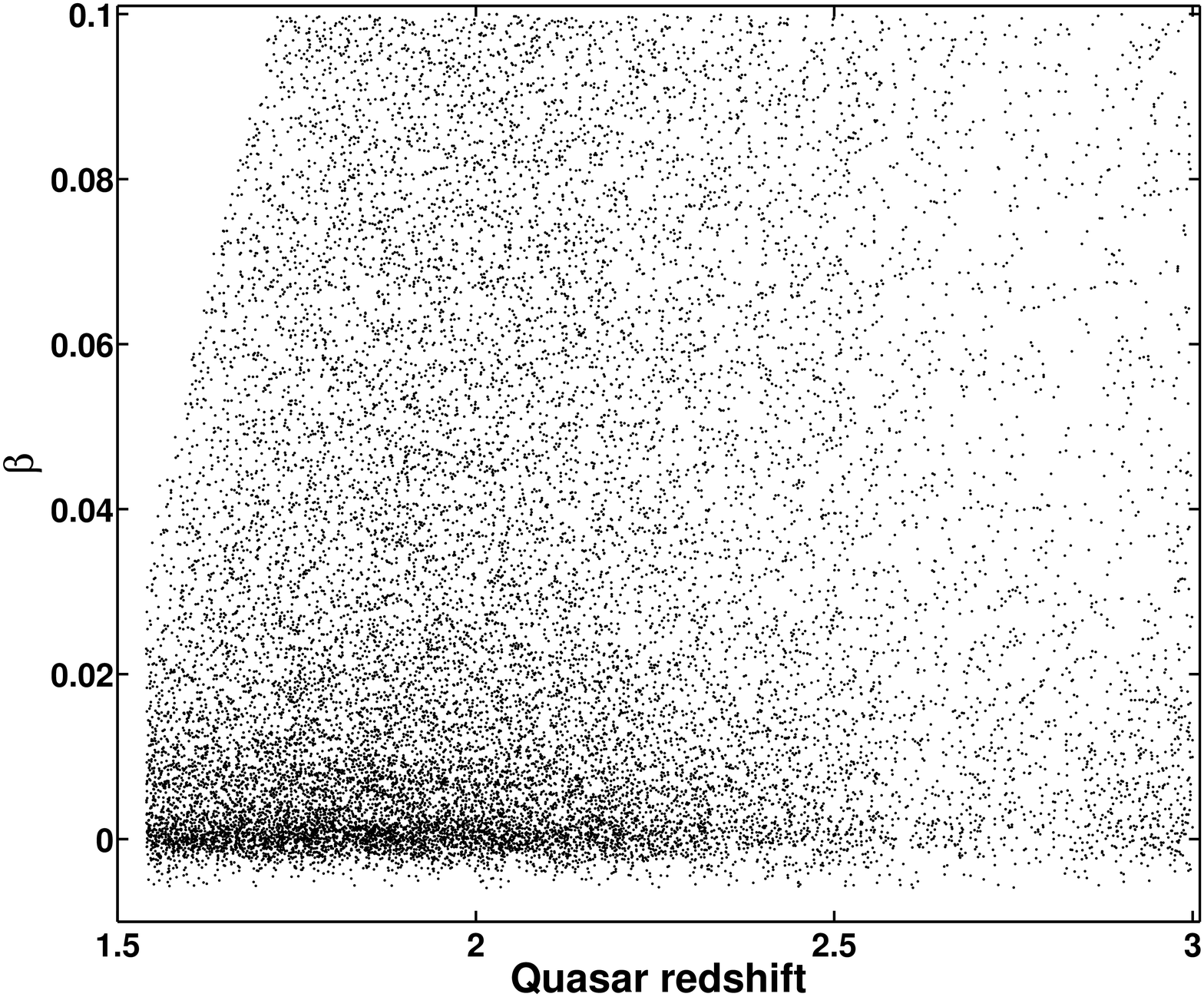}
\caption{The 18\,908 \civ absorbers in the non-BALQSO sample plotted
as a function of the quasar redshift and outflow velocity,
$\beta$. The absence of absorbers at large $\beta$ for quasar
redshifts $z$$<$1.75 results from the minimum observed-frame
wavelength of $\simeq$3810\,\AA \ included in the SDSS spectra.}
 \label{f:qz_beta}
\end{figure}

\subsection{Absorber outflow velocities}

The outflow/inflow velocity of a \civ absorber, detected at wavelength
$\lambda_{abs}$ in a rest-frame quasar spectrum, is parametrized by
$\beta = v/c \simeq (1548.19 - \lambda_{abs})/1548.19$, where $c$ is
the velocity of light and 1548.19\,\AA \ is the rest-frame wavelength
of the blue component of the \civ doublet. An outflowing absorber has
positive $\beta$ and the \civ absorber search was undertaken for
$\beta\le0.1$, i.e. to $v\simeq$30\,000\kms. At larger $\beta$, the
presence of absorption due to \siiv$\lambda \lambda$ 1393.8,1402.8 increases the
number of spurious \civ absorber detections, particularly in the
BALQSO sample. The significantly improved errors in the quasar
redshifts \citep{2010MNRAS.405.2302H} reduces the number of apparently
infalling absorbers (with negative $\beta$) dramatically. Absorbers
are still, however, detected with negative $\beta$ due to i) remaining
quasar redshift errors, ii) the presence of `associated' \civ
absorbers that are not participating in an outflow, and iii)
[potentially] infalling \civ absorbers. The distribution of 18\,908
\civ absorbers in the non-BALQSO sample as a function of $\beta$ is
shown in Fig.~\ref{f:betadist} and the distribution as a function of
the quasar redshift and $\beta$ is shown in Fig.~\ref{f:qz_beta}.
Throughout the paper, the statistics given for absorbers apply to
those with $\beta \ge 0$, unless otherwise stated.

\citet{2008MNRAS.386.2055N} drew attention to the shape of the \civ
absorber distribution at small $\beta$ (their figs. 5 and 6). The
twin-peak, or, equivalently, reduced number of outflow absorbers at
$v$$<$2000\kms ($\beta$$<$$0.00667$) is confirmed at high significance in
Fig.~\ref{f:betadist}. The apparent two-component nature of the
absorber velocity distribution is perhaps most evident in
Fig.~\ref{f:qz_beta}, where the much larger number of absorbers and
the effect of the improved redshift accuracy for the quasars, confirms
the presence of a significant absorber population centred on the
systemic quasar redshift and the presence of an apparent maximum in
the outflow absorber population at $\simeq$0.00667. Analysis of the
different absorber populations is beyond the scope of the results
presented here, but further investigation of \civ absorbers employing
the significant statistical power of the new SDSS DR10 quasar sample
\citep{2014A&A...563A..54P} will be forthcoming.

\begin{figure*}
\includegraphics[width=188mm]{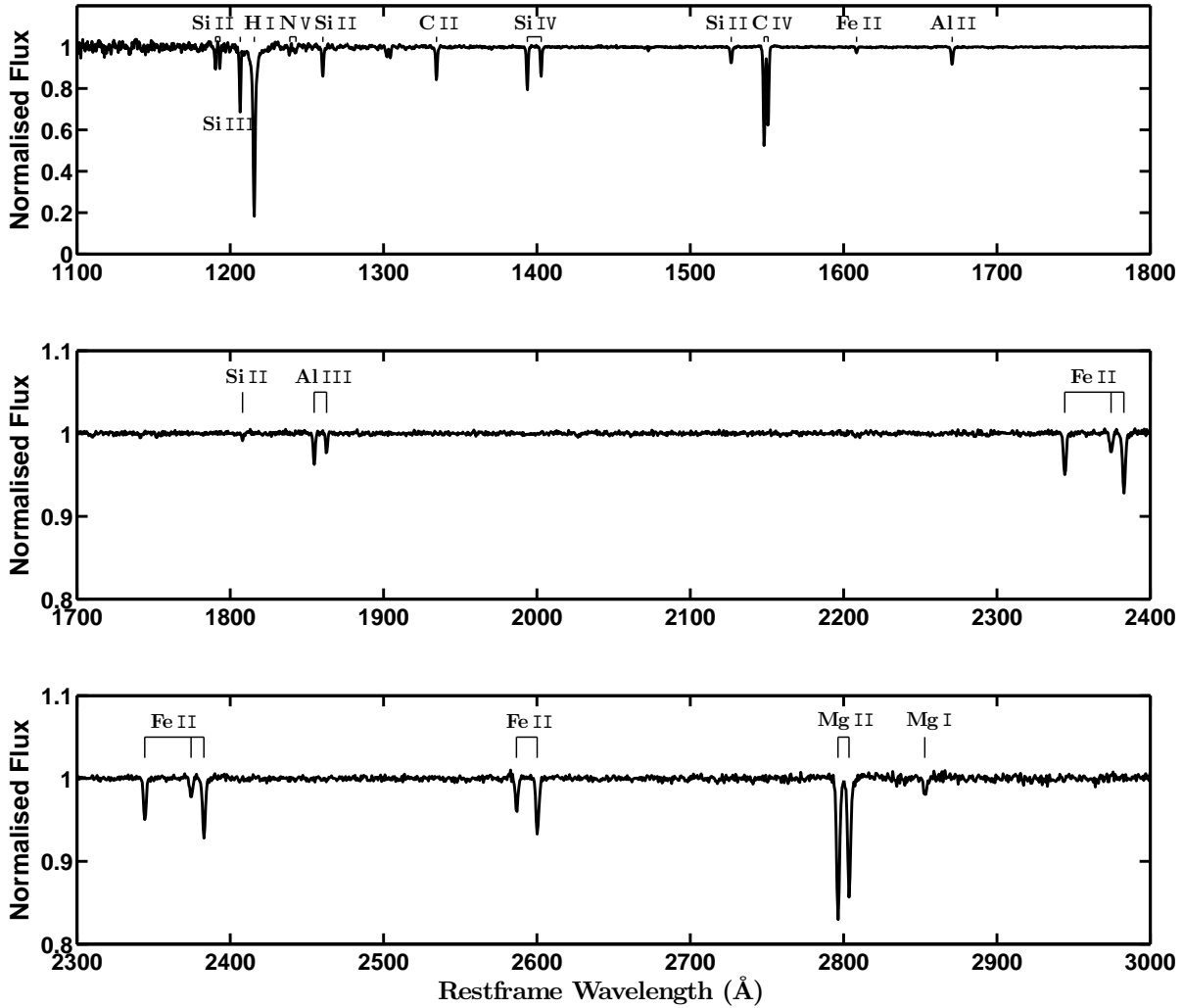}
\caption{Arithmetic-mean composite absorption-line spectrum of 5098
\civ absorbers with $0.04 < \beta < 0.1$ identified in the non-BALQSO
sample. The individual quasars containing the absorbers were
normalised using a high S/N quasar composite
(Section~\ref{sec:ll_prop}) and then combined in the rest-frame of the
absorbers. The number of absorbers contributing to the composite
decreases towards the wavelength extremes, causing the reduction in
the S/N, but always exceeds 300 absorbers. The location of strong
absorption lines are indicated, including features evident in
later figures, e.g. the \feii\l1608 and \alii\l1671 lines (see
Fig.~\ref{f:pixlag}). Note the change in the y-axis range for the top
panel.}
\label{f:abs_comp}
\end{figure*}

Our absorption line detection routine resulted in samples of 16\,762
(2637) \civ doublets ($0.0 \le \beta \le 0.1$) from the 31\,142 (2654)
non-BALQSO (BAL) spectra respectively. In Fig.~\ref{f:abs_comp} we
show the arithmetic-mean composite absorption-line spectrum of 5098
\civ absorbers with $0.04 < \beta < 0.1$ identified in the non-BALQSO
sample. The individual quasars containing the absorbers were
normalised using a high-S/N quasar composite (see
Section~\ref{sec:ll_prop} for details) and combined, after shifting
the normalised-spectra to the rest-frame of the absorbers. Numerous
additional absorption features from other ions associated with the
\civ systems, covering a range of ionization potential, are visible in
the high-S/N composite spectrum.

\subsection{Identifying triplet \civ absorbers}\label{ss:trip_ident}

With the \civns-absorber sample in hand it is straightforward to
search for potentially line-locked \civ systems via the distribution
of separations from \civns-doublets to other (individual) absorbers
blueward of the \civ doublet. The presence of line-locked \civ systems
would lead to an excess of separations of absorbers at $\simeq$14.5
pixels, indicating the presence of three, equally-spaced, absorption
features, henceforth `triples'.

Figure~\ref{f:pixlag} shows the distribution of \civns-doublet to
absorber separations for \civ doublets with $0.05 < \beta < 0.10$
identified in the non-BALQSO sample. The separation is measured from
the position of the blue (1548.19\,\AA) line in each \civ doublet.
The distribution is dominated by the 4579 pairs with separations of
7.24$\pm$1.00 pixels; the \civ doublets themselves. At larger
pixel-lags $\simeq$165 and $\simeq$332\,pixels), the detection of
\feii\l1608.5 and \alii\l1670.8 absorption respectively, associated
with some of the \civ absorbers (see Fig.~\ref{f:abs_comp})
demonstrates the effectiveness of the absorption-line identification
scheme. Evidence for some line-locked \civ systems can be seen from
the small, but significant, excess of $\simeq$87$\pm$13 absorbers
around $\simeq$14\,pixels. The exact number of doublet-absorber pairs
at larger separations depends primarily on the velocity-separations of
the \civ doublets from the bulk of the outflow-related absorbers
present at $\beta$$\la$0.02 in the quasars (Fig.~\ref{f:betadist}).
The broad excess of doublet-absorber detections centred at $\sim$250
pixels results from the increased number of such pairs.

Considering the number of \civ doublet to absorber separations for
doublets at smaller values of $\beta$ reveals a clear detection of an
excess of absorbers at $\simeq$14.5\,pixels, indicating the potential
presence of line-locked \civ systems in both the non-BALQSO and
BALQSO samples. Figure~\ref{f:pixlag_2} shows the histograms of \civ
doublet to absorber separations for doublets with
$0.0$$<$$\beta$$<0.05$ in the 31\,142 non-BALQSO and 2654 BALQSO
samples.  Excesses of $\simeq$800 (220) absorber-pairs at $\simeq$14.5
pixels for the non-BALQSO (BALQSO) samples are present. The number of
\civ doublets is much smaller in the BALQSO sample but no other
significant excess of doublet-absorber pairs is evident at other
velocity separations in either of the samples.

\section{The statistics of line-locked systems}\label{sec:ll_stats}

\subsection{\civ absorbers of different physical
  origin}\label{ssec:ll_stats_orig}

The absorber-identification scheme described in the previous section,
while effective, is essentially blind to the intrinsic nature of the
absorber systems found. In addition to the outflowing absorbers that
are the focus of the investigation, there are four additional sources
of \civ doublets in the absorber catalogue: (a) spurious systems due
to the limited S/N of the quasar spectra and hence of the absorber
detections, (b) apparent \civ systems resulting from the presence of
one, or two, real narrow Gaussian-like absorption features arising
from species other than \civns, (c) real \civ absorbers due to
intervening gaseous systems that are not directly related to the
quasars but possess redshifts that apparently place them within
30\,000\kms \ of the quasar redshift, (d) real \civ absorbers
resulting from gaseous systems that are directly related to the
presence of the quasar, e.g. within the quasar host-galaxy halo or
associated with galaxies in the same group or cluster in which the
quasar resides. The incidence of line-locking of interest is that
relative to the proportion of \civ systems in outflows and a
statistical correction to the observed number of absorbers is
desirable for each of the four `contaminant' populations.

\begin{figure}
\includegraphics[width=84mm]{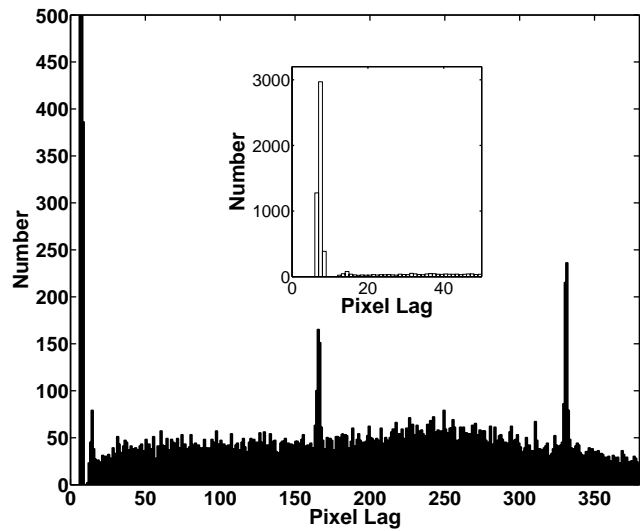}
\caption{A histogram of the separation between the blueward trough of
an identified \civ doublet, and all other identified individual
absorbers in that quasar spectrum, for \civ doublets with $0.05 \le
\beta \le 0.10$ (i.e. at velocities of $v_{outflow}$$>$15\,000\kms)
identified in the 31\,142 non-BALQSO sample. The distribution is
dominated by the 4579 detections of the \l1550.5\,\AA \ line in
each doublet at 7.24$\pm$1.00 pixels --- i.e. the \civ doublets
themselves (shown in the smaller enclosed plot). At larger pixel-lags,
($\simeq$165 and 332\,pixels), the detection of \feii\l~1608.5 and
\alii\l~1670.8 absorption respectively, associated with some of the
\civ doublets (see Fig.~\ref{f:abs_comp}), demonstrates the
effectiveness of the absorption-line identification scheme. Evidence
for the presence of some line-locked \civ systems can be seen from the
small, but significant, excess of $\simeq$90 absorbers around
$\simeq$14\,pixels.}
 \label{f:pixlag}
\end{figure}

The number of apparent \civ doublets, due to the coincidence of two
spurious Gaussian absorbers with the velocity separation of a \civ
doublet, was estimated by performing a search for Gaussian `emission'
features in the same quasar spectra (both the non-BALQSO and BALQSO
samples). From the resulting catalogues of low S/N emission features,
a total of 36$\pm$2.5 and 16$\pm$1.5 spurious \civ doublets were
`identified' per 0.01 interval in $\beta$ for the non-BALQSO and
BALQSO samples respectively. The expected number of \civ
triple-systems resulting solely from the coincidence of three spurious
absorber detections is essentially zero.

The majority of both \civ doublets and triplets that are not real
arise from the coincidence of genuine, usually weak, absorption
features present at the velocity separations used to define \civ
doublets or triplets.  The probability of finding a narrow
Gaussian-like absorber at a given velocity separation from a location
chosen at random within the quasar spectra is straightforward to
determine empirically. Specifically, on average the probability is
equal to 0.52 (0.76) per cent, per pixel, at small velocity
separations ($\la$1500\kms) for the non-BALQSOs (BALQSOs). Taking the
sample of Gaussian absorbers in the non-BALQSO sample, there is
therefore an approximately 1/200 chance that another Gaussian absorber
will be found in a single pixel at a specified velocity separation. A
\civ doublet was defined by requiring the presence of a second
Gaussian absorber within $\pm$1.0\,pixels of the predicted 500\kms
doublet separation and thus 1.04 (1.53) per cent of the \civ doublets
are predicted to be spurious. It is worth noting that the
contamination of the sample of \civ doublets due to effects arising
from both the finite S/N of the spectra and the presence of unrelated
absorbers (i.e. correction factors `a' and `b' above) is thus very
small.

Triplet absorbers were defined by requiring the presence of a third
Gaussian absorber within $\pm$1.5\,pixels of the predicted location of
a line-locked absorber associated with each \civ doublet\footnote{The
$\pm$1.5\,pixel definition was determined from the observed
velocity-width evident for absorbers associated with the \civ
doublets, e.g. the \feii\l1608 and \alii\l1671 lines in
Fig~\ref{f:abs_comp}, much of which results from the low S/N of the
individual absorber detections. Other than the exact number of systems
defined, none of the results in the paper are sensitive to the precise
definition of doublets or triplets in velocity space.}. The number of
spurious triples is therefore equal to 1.52 (2.29) per cent {\it of
the number of \civ doublets} identified in the non-BALQSO (BALQSO)
samples. As a consequence, the correction to the observed number of
triplet absorbers is fractionally quite large, particularly at small
values of $\beta$, where the number of outflow \civ doublets is
significant.

Potential higher-order systems, consisting of four or more absorbers
separated by 7.24-pixels, were also sought. The number of spurious
systems includes a component (exactly analogous to the calculation for
the triplets) equal to 1.52 (2.29) per cent of the number of \civ
triplets identified in the non-BALQSO (BALQSO) samples. In addition,
the probability that two real \civ doublets, separated by 14.5-pixels,
occur is not insignificant. Again, the number of such instances
expected is readily calculated from the statistics of the number of
\civ doublets identified per $\beta$-interval.

The number of intervening \civ absorbers present in the absorber
catalogue was calculated using the results from
\citet{2013ApJ...763...37C}. For the EW$_{1548}$-interval $0.4 < {\rm
EW}_{1548} < 0.9$ \AA, where the absorber sample is close to complete,
the slope of the power-law model for the EW frequency distribution
($f({\rm EW}_{1548})$) of \citet{2013ApJ...763...37C}, $f({\rm
EW}_{1548})=k e^{\alpha {\rm EW_{1548}}}$, with $k$=3.49 and
$\alpha$=-2.62, is in excellent agreement with the observed $f({\rm
EW}_{1548})$ in the non-BALQSO sample.  Calculating the redshift-path
for detection of \civ absorbers in the 31\,142 quasars of the
non-BALQSO sample as a function of $\beta$ produces a predicted number
of intervening \civ absorbers. For the $\beta$-interval $0.05 \le
\beta \le 0.10$, the \citet{2013ApJ...763...37C} model predicts
1329$\pm$140 intervening absorbers, compared to the total of 2063
absorbers detected. Thus, 64.4$\pm$8\,per cent of the detected
absorbers with $0.05 \le \beta \le 0.10$ are estimated to be
intervening.

A correction for the number of intervening \civ absorbers for the full
sample was made by assuming that the observed excess of associated
systems in the $0.05< \beta \le 0.1$ interval is independent of
absorber rest-frame EW$_{1548}$; leading to a predicted number of
intervening absorbers [in the full EW$_{1548}$-range] of 723$\pm$58
(62$\pm$5) per 0.01-interval in $\beta$ for the entire non-BALQSO
(BALQSO) sample.  The EW$_{1548}$-extrapolation involved is small
($\simeq$80\,per cent of the entire absorber sample possess
EW$_{1548}$$\ge$0.3\,\AA) and the key results reported below are
insensitive to the exact level of the correction for intervening
absorbers because the sample is dominated by outflowing absorbers at
$\beta$$<$0.05.

The same correction for the presence of intervening absorbers was
applied to the BALQSO sample, taking into account the smaller total
redshift path.

\begin{figure*}
\includegraphics[width=188mm]{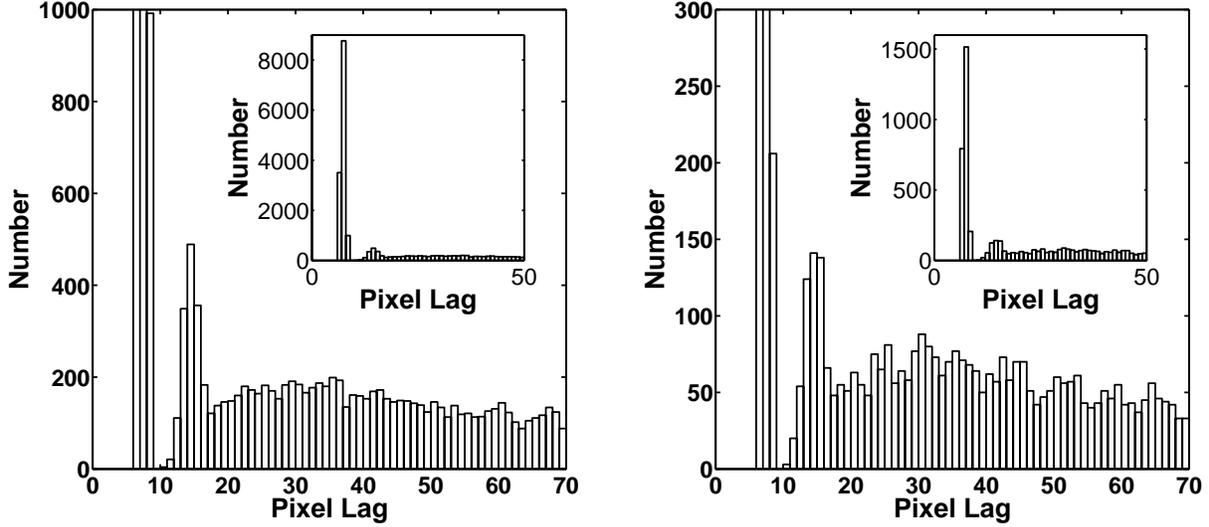}
\caption{As for Fig.~\ref{f:pixlag}, histograms of the separation
between the blueward trough of an identified \civ doublet, and all
other identified individual absorbers in that quasar spectrum.
Separations for absorbers with velocity $0.00 \le \beta
\le 0.05$ identified in the 31\,142 non-BALQSOs (left panel) and in the
2654 BALQSOs (right panel) are shown. The distribution is dominated by the
$\simeq$12\,000 (1800) detections of the \l1550.5\,\AA \ line in
each doublet at 7.24$\pm$1.00 pixels for the non-BALQSOs (BALQSOs)---
(shown in the smaller enclosed plots). The excess of $\simeq$800 (220)
absorber-pairs centred around 14.5 pixels provides clear evidence for
a population of \civ triplet absorbers in the non-BALQSO (BALQSO)
sample.}
 \label{f:pixlag_2}
\end{figure*}

At small $\beta$ values ($\beta$$<$0.01) there is evidence
(Fig.~\ref{f:betadist}) for a population of `associated'-absorbers
(`d' above) that do not participate in any outflow present. Without
additional information it is not straightforward to derive an accurate
estimate of the number of such associated-absorbers as a function of
$\beta$. The exclusion of the $\beta$$<$0.00 absorbers results in a
very crude, partial, correction by removing absorbers that, given the
level of the quasar redshift uncertainties, are unlikely to form
part of any outflow. The statistics of the line-locked systems at
small $\beta$ values thus represent lower-limits to the fraction of
such systems present in any outflows.

\subsection{The fraction of line-locked \civ absorbers}

Tables \ref{tab:lstat_nbal} (non-BALQSOs) and \ref{tab:lstat_bal}
(BALQSOs) include the observed and corrected numbers of \civ doublets
and triplets as a function of $\beta$.  The figures in Tables
\ref{tab:lstat_nbal} and \ref{tab:lstat_bal} for $\beta$$\le$0.03
demonstrate unambiguously that approximately 10\,per cent of \civ
absorbers associated with quasar outflows are line-locked. Given the
very small number of individual line-locked absorber systems reported
in the literature the result was not anticipated. Also of potential
interest is the presence of line-locked narrow \civ absorbers in the
BALQSO sample, with a frequency very similar to that in the non-BALQSO
sample.

\begin{table*}
\caption{\civ doublet and triplet systems in the non-BALQSO sample.
Column 1 specifies the range in $\beta$ and Column 2 gives the
observed number of \civ doublets.  The number of \civ doublets after
correction for the number of spurious systems (i.e. factors `a' and
`b' described in the text) is given in Column 3 and the number of
\civ doublets after subtraction of the number of
intervening systems, unrelated to the quasar (i.e. including factor
`c' described in the text) is given in Column 4.  Columns 5 and 6
provide the observed and corrected number of \civ triplets (there are
no intervening systems to be subtracted). Finally, Columns 7 and 8
provide the observed and corrected numbers of higher-order multiple
\civ absorbers, i.e. quartets, quintets,...). The number of multiple
absorbers with five or more components is small and the distribution
is completely dominated by the number of quadruple absorbers.
}
\label{tab:lstat_nbal}
\centering
%\begin{scriptsize}
\begin{tabular}{crrrrrrr}\\ 
\hline
\multirow{1}{*}{} & \multicolumn{3}{c}{Doublets} &
\multicolumn{2}{c}{Triplets} &
\multicolumn{2}{c}{Quadruplets+}\\

\multicolumn{1}{c}{$\beta$} & 
\multicolumn{1}{c}{$N_{Obs}$} &
\multicolumn{1}{c}{$N_{Corr(ab)}$} &
\multicolumn{1}{c}{$N_{Corr(abc)}$} &
\multicolumn{1}{c}{$N_{Obs}$} &
\multicolumn{1}{c}{$N_{Corr(ab)}$} &
\multicolumn{1}{c}{$N_{Obs}$} &
\multicolumn{1}{c}{$N_{Corr(ab)}$} \\
\hline
0.00-0.01   & 4968$\pm$71 & 4873$\pm$71 & 4150$\pm$71 &  574$\pm$24 & 486$\pm$24 & 90$\pm$10 & 41$\pm$10 \\
0.01-0.02   & 2301$\pm$48 & 2239$\pm$48 & 1516$\pm$48 &  163$\pm$13 & 124$\pm$13 & 41$\pm$7 & 27$\pm$7 \\
0.02-0.03   & 1573$\pm$40 & 1520$\pm$40 &  797$\pm$40 &   73$\pm$9 &  47$\pm$9 &   6$\pm$3 &   1$\pm$3 \\
0.03-0.04   & 1230$\pm$35 & 1182$\pm$35 &  469$\pm$35 &   41$\pm$6 &  21$\pm$6 &   4$\pm$2 &   0$\pm$2 \\
0.04-0.05   & 1093$\pm$33 & 1047$\pm$33 &  371$\pm$33 &   22$\pm$5 &   5$\pm$5 &   4$\pm$2 &   2$\pm$2 \\
0.050-0.075 & 2321$\pm$48 & 2219$\pm$48 &  666$\pm$48 &   68$\pm$8 &  31$\pm$8 &   5$\pm$2 &  -1$\pm$2 \\
0.075-0.100 & 2111$\pm$46 & 2030$\pm$46 &  864$\pm$46 &   65$\pm$8 &  31$\pm$8 &   8$\pm$3 &   3$\pm$3 \\
\end{tabular}
\begin{minipage}{165mm}
\end{minipage}
%\end{scriptsize}
\end{table*}

\begin{table*}
\caption{\civ doublet and triplet systems in the BALQSO sample. Column
definitions are as for Table~\ref{tab:lstat_nbal}}
\label{tab:lstat_bal}
\centering
%\begin{scriptsize}
\begin{tabular}{crrrrrrr}\\ 
\hline
\multirow{1}{*}{} & \multicolumn{3}{c}{Doublets} &
\multicolumn{2}{c}{Triplets} &
\multicolumn{2}{c}{Quadruplets+}\\

\multicolumn{1}{c}{$\beta$} & 
\multicolumn{1}{c}{$N_{Obs}$} &
\multicolumn{1}{c}{$N_{Corr(ab)}$} &
\multicolumn{1}{c}{$N_{Corr(abc)}$} &
\multicolumn{1}{c}{$N_{Obs}$} &
\multicolumn{1}{c}{$N_{Corr(ab)}$} &
\multicolumn{1}{c}{$N_{Obs}$} &
\multicolumn{1}{c}{$N_{Corr(ab)}$} \\
\hline 
0.00-0.01   & 603$\pm$25 & 576$\pm$25 & 514$\pm$25 & 102$\pm$10 & 85$\pm$10 & 25$\pm$5 & 14$\pm$5 \\
0.01-0.02   & 511$\pm$23 & 485$\pm$23 & 423$\pm$23 & 101$\pm$10 & 87$\pm$10 & 21$\pm$5 & 11$\pm$5 \\
0.02-0.03   & 344$\pm$19 & 322$\pm$19 & 260$\pm$19 &  46$\pm$7 & 37$\pm$7 &  15$\pm$4 &   9$\pm$4 \\
0.03-0.04   & 217$\pm$15 & 197$\pm$15 & 136$\pm$15 &  13$\pm$4 &  8$\pm$4 &   3$\pm$2 &   2$\pm$2 \\
0.04-0.05   & 148$\pm$12 & 131$\pm$12 &  73$\pm$12 &  11$\pm$3 &  7$\pm$3 &   1$\pm$1 &   1$\pm$1 \\
0.050-0.075 & 254$\pm$16 & 216$\pm$16 &  84$\pm$16 &  18$\pm$4 & 10$\pm$4 &   0$\pm$1 &   0$\pm$1 \\
0.075-0.100 & 189$\pm$14 & 160$\pm$14 &  61$\pm$14 &  13$\pm$4 &  8$\pm$4 &   2$\pm$1 &   2$\pm$1 \\
\end{tabular}
\begin{minipage}{165mm}
\end{minipage}
%\end{scriptsize}
\end{table*}

The observed number of quadruple systems, potentially representing
three line-locked \civ doublets, is small. The majority are predicted
to be due to chance occurrences of [non-line-locked] doublets, a
conclusion that is confirmed via construction of composite absorber
spectra (Section~\ref{s:res_qseds_noll}). The reality of the
line-locked triplet systems is established explicitly in
Section~\ref{sec:ll_prop}, but first the prevalence of line-locked
absorbers in the subset of quasars where outflowing triplets can be
detected is considered.

\subsection{The fraction of quasar outflows with line-locked \civ
  absorbers}

Line-locked \civ absorbers can only be detected in quasars where two
or more \civ absorbers in an outflow are present. Given the frequency
of \civ absorbers in the quasar samples, the majority of quasars
possess only a single \civ absorber, which, in many cases, is due to
an unrelated intervening absorber not associated with an
outflow. Consideration of the statistics of line-locked systems in
quasars with two or more \civ absorbers is
revealing. Tables~\ref{tab:qstat_nbal} and \ref{tab:qstat_bal} present
the number of absorbers present in the non-BALQSO and BALQSO samples,
excluding quasars with only a single \civ doublet detected.

\begin{figure*}
\includegraphics[width=188mm]{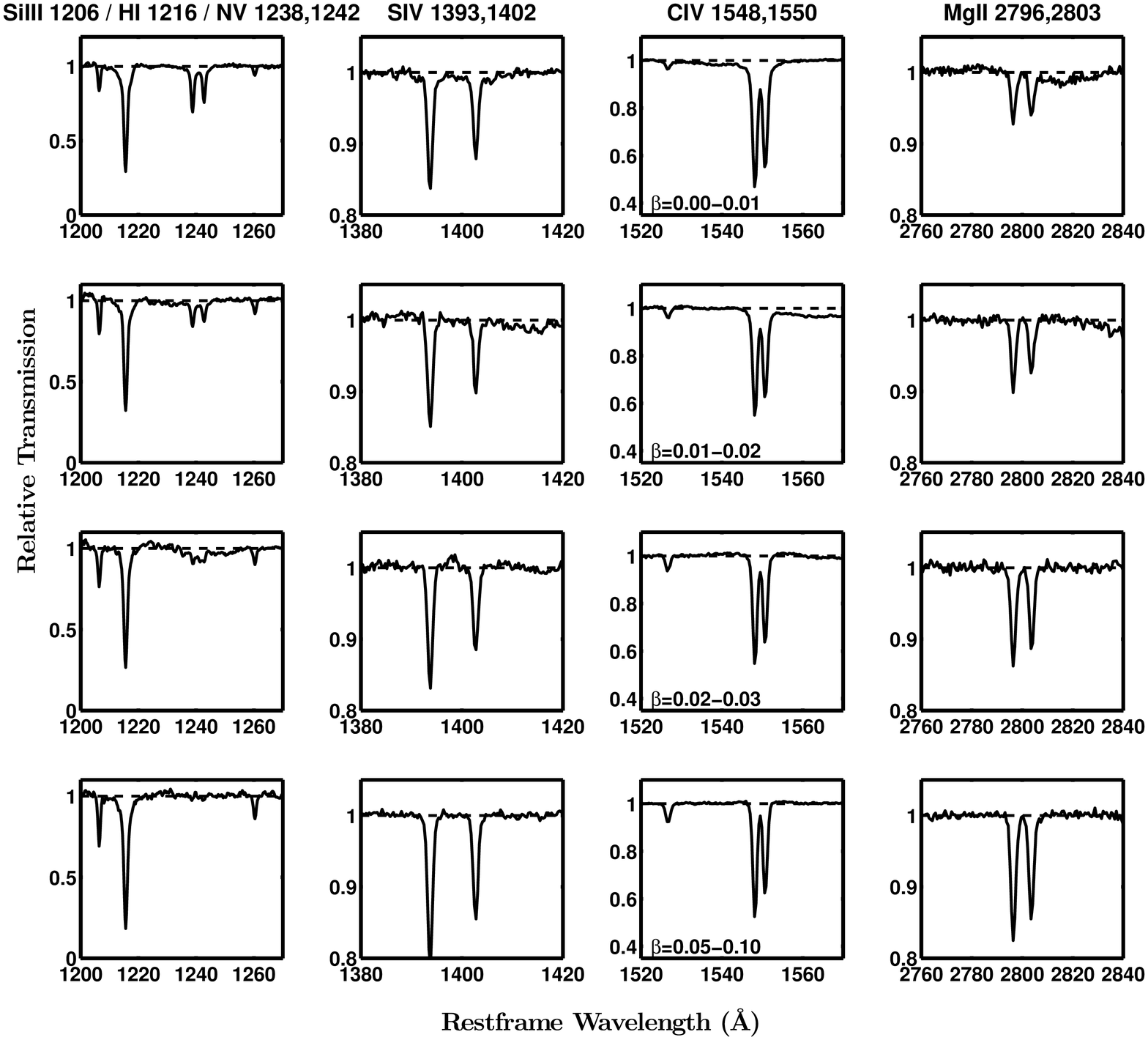}
 \caption{Arithmetic-mean composite absorption-line spectra of \civ
doublets in quasars with only a single absorber. The signature of the
quasar SED has been removed via the use of a composite quasar SED, as
described in (Section~\ref{sec:ll_prop}). From left to right, the
panels show four wavelength intervals containing strong absorption
features. From top to bottom, the panels show the absorption-line
spectra in intervals of increasing $\beta$, as specified in the panels
of the third column. A strong systematic trend is present as a function of
increasing $\beta$, with lower ionization species (e.g. \mgiins)
strengthening and higher ionization species (e.g. \nvns) weakening.
}
 \label{f:comp_ds}
\end{figure*}

\begin{table*}
\caption{Triplet systems in the non-BALQSO quasars with multiple \civ
absorbers.  The content of Tables \ref{tab:qstat_nbal} and
\ref{tab:qstat_bal} is as follows. Column 1; range in $\beta$. Column
2; number of quasars with multiple \civ absorbers. Column 3; total
number of \civ doublets and triplet systems.  Column 4; corrected
number of quasars with multiple \civ absorbers after statistically
removing pairs of intervening \civ doublets and pairs of intervening
and outflow doublets. Column 5; the number of triplet-systems alone.
Column 6; corrected number of triplet absorbers after allowing for the
number of spurious systems (using the same correction scheme as in
columns 5 and 6 of Tables~\ref{tab:lstat_nbal} and
\ref{tab:lstat_bal}. Column 7; percentage of quasars with multiple
absorbers due to outflowing material that possess line-locked \civ
absorbers.
}
\label{tab:qstat_nbal}
\centering
%\begin{scriptsize}
\begin{tabular}{crrrrrr}\\ 
\hline
\multicolumn{1}{c}{$\beta$} & 
\multicolumn{1}{c}{$N_{abs}$} &
\multicolumn{1}{c}{$N_{quasar}$} &
\multicolumn{1}{c}{$NC_{quasar}$} &
\multicolumn{1}{c}{$N_{Trip}$} &
\multicolumn{1}{c}{$NC_{Trip}$} &
\multicolumn{1}{c}{$F_{Trip}$} \\
\hline
0.00-0.01 & 1115 &  819 & 617 & 572 & 484 & 78$\pm$5\% \\
0.00-0.02 & 2299 & 1413 & 973 & 729 & 603 & 62$\pm$4\% \\
0.00-0.03 & 3240 & 1850 & 1124 & 802 & 651 & 58$\pm$4\% \\
0.00-0.04 & 3966 & 2165 & 1116 & 837 & 667 & 60$\pm$4\% \\
0.00-0.05 & 4669 & 2460 & 1047 & 858 & 671 & 64$\pm$4\% \\
0.0067-0.0233 & 939 & 607 & 393 & 328 & 276 & 70$\pm6\%$ \\
\end{tabular}
\begin{minipage}{165mm}
\end{minipage}
%\end{scriptsize}
\end{table*}

\begin{figure*}
\includegraphics[width=188mm]{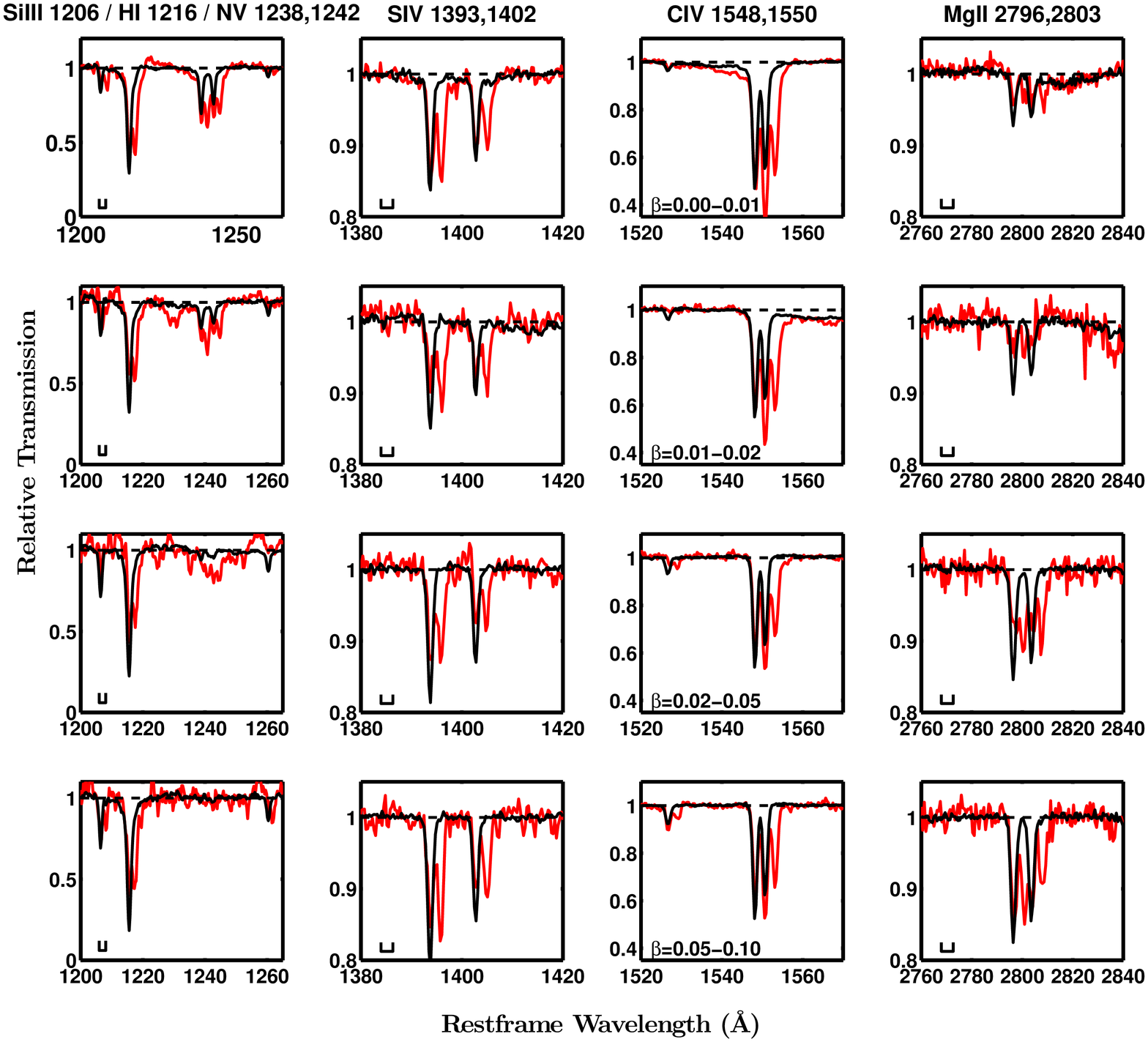}
 \caption{Arithmetic-mean composite absorption-line spectra of \civ
triplet absorbers (red) with the composite absorption-line spectra of
\civ doublets in the non-BALQSO sample with only a single absorber per
quasar (c.f. Fig.~\ref{f:comp_ds}) overplotted (black). From left to
right, the panels show four wavelength intervals from each
composite. From top to bottom, the panels show the absorption-line
spectra in intervals of increasing $\beta$.  The presence of `double'
\hins, \nvns, \siiv and \mgii absorbers separated by the 500\kms
interval (indicated by the lines in the bottom lefthand corner of each
panel) resulting from two line-locked \civ systems demonstrates
unambiguously the reality of the line-locked systems. The only
slightly weaker EWs of the second absorber systems at redder
wavelengths, (e.g. in \siiv and \mgiins) confirms that the contamination
of the \civ triplet sample due to spurious absorbers is small. The
same systematic trend for species with different ionization potentials
as a function of $\beta$ is present for the line-locked systems and
the \civ doublets.}
 \label{f:comp_tds}
\end{figure*}

\begin{table*}
\caption{Triplet systems in the BALQSO quasars with multiple
\civ absorbers. Column definitions are as for Table~\ref{tab:qstat_nbal}.}
\label{tab:qstat_bal}
\centering
%\begin{scriptsize}
\begin{tabular}{crrrrrrrr}\\ 
\hline
\multicolumn{1}{c}{$\beta$} & 
\multicolumn{1}{c}{$N_{abs}$} &
\multicolumn{1}{c}{$N_{quasar}$} &
\multicolumn{1}{c}{$NC_{quasar}$} &
\multicolumn{1}{c}{$N_{Trip}$} &
\multicolumn{1}{c}{$NC_{Trip}$} &
\multicolumn{1}{c}{$F_{Trip}$} \\
\hline
0.00-0.01   &  202 & 149 & 118 & 101 &  84 & 71$\pm$11\% \\
0.00-0.02   &  647 & 382 & 309 & 194 & 164 & 53$\pm$7\% \\
0.00-0.03   &  988 & 518 & 403 & 234 & 196 & 49$\pm$6\% \\
0.00-0.04   & 1189 & 591 & 430 & 245 & 203 & 50$\pm$6\% \\
0.00-0.05   & 1344 & 638 & 425 & 256 & 210 & 49$\pm$6\% \\
0.0067-0.0233 & 431 & 269 & 216 & 161 & 135 & 63$\pm8\%$ \\

\end{tabular}
\begin{minipage}{165mm}
\end{minipage}
%\end{scriptsize}
\end{table*}

\begin{figure*}
\includegraphics[width=188mm]{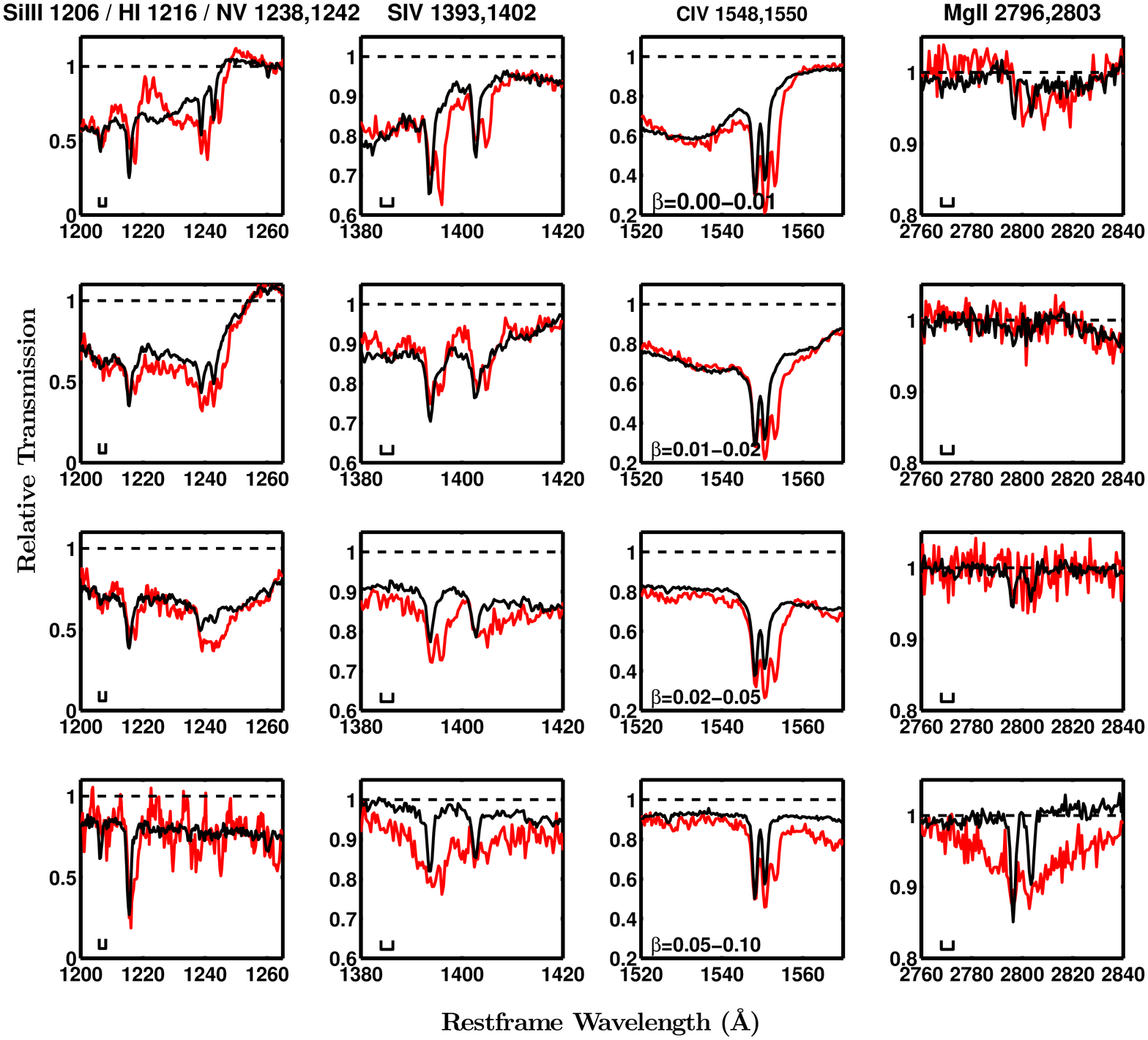}
\caption{Arithmetic-mean composite absorption-line spectra of \civ
triplet absorbers (red) with the composite absorption-line spectra of
\civ doublets in the BALQSO sample with only a single absorber per
quasar overplotted (black). From left to right, the panels show four
wavelength intervals from each composite. From top to bottom, the
panels show the absorption-line spectra in intervals of increasing
$\beta$.  The much smaller number of quasar spectra in the BALQSO
sample means that the S/N of the composites is significantly poorer
than for the non-BALQSO sample, particularly for \mgii and at large
$\beta$, where the number of absorbers is small. The presence of
`double' \hins, \nv and \siiv absorbers separated by the 500\kms~interval (indicated by the lines in the bottom lefthand corner of each
panel) resulting from two line-locked \civ systems again demonstrates
unambiguously the reality of the line-locked systems.  The large-scale
depression of the spectra, away from unit transmission, results from
the presence of the extended BAL-troughs in the quasars [that are not
present in the control-sample employed to remove the signature of the
quasar SED]. The same systematic trend for species with different
ionization potentials as a function of $\beta$, evident in the
non-BALQSO sample, is present.
}
 \label{f:comp_tdsbals}
\end{figure*}

To investigate the frequency of line-locked systems it is necessary to
define a velocity (i.e. $\beta$) interval in which two or more
absorbers are detected, which in turn requires some knowledge of the
extent of an outflow. In the limit of the velocity interval
$\Delta\beta$$\rightarrow$0, only line-locked systems (i.e. `triples')
will be included and the line-locked frequency will be
$\simeq$100\,per cent. At the other extreme, using a very large
velocity interval will lead to a sample dominated entirely by
intervening doublet absorbers. Given the distribution of \civ
absorbers as a function of $\beta$ (Fig.~\ref{f:betadist}), outflow
extents of order 3000$-$10\,000\kms~are an appropriate
choice. Examining the $\beta$-range $\beta$$\le$0.05 ensures that a
significant fraction of the absorbers are associated with outflows.

For completeness, Tables~\ref{tab:qstat_nbal} and \ref{tab:qstat_bal}
include the statistics for $\beta$-intervals of 0.01 to 0.05, starting
at $\beta$=0.0, along with the results for a 5000\kms interval
starting at $\beta$=0.0067 (i.e. 2000\kms), chosen to focus on an
outflow velocity range, some ten times the \civ line-locking scale
(i.e. 500\kms), that avoids the majority of the `associated'-absorbers
[at small $\beta$] while minimising the fraction of intervening \civ
absorbers. 

Uncorrected and corrected figures for the number of quasars with
multiple \civ absorbers and the number with line-locked triplet
absorbers are included in the tables. The correction for the number of
spurious triplets in each $\beta$-interval is straightforward to
calculate using the numbers from columns 5 and 6 of
Tables~\ref{tab:lstat_nbal} and \ref{tab:lstat_bal}. The reduction in
the number of quasars with multiple absorbers due to removal of the
fraction of spurious triples is less than the loss of triples because
some quasars possess two \civ doublets and a triplet.  A statistical
correction to the number of quasars with multiple absorbers due to the
occurrence of a) two intervening \civ doublets and b) an intervening
\civ doublet and a single \civ doublet in an outflow has also been
applied. The probability that a quasar possesses an intervening
absorber is known to high accuracy for each $\beta$-interval and the
number of quasars with pairs of such absorbers is calculated assuming
the absorbers are independent (i.e. there is no clustering; which,
in practice, is a good approximation).

The number of \civ doublets resulting from outflows in each
$\beta$-interval is also known to high accuracy. The correction for
the presence of pairs of intervening and outflow \civ doublets is
calculated assuming that both the intervening and outflow doublets
are distributed randomly. The latter assumption is not exactly correct
as at least some outflows will produce multiple \civ doublets in
individual quasars. However, the distribution of outflow doublets,
for all the $\beta$ intervals, among the 31\,142 non-BALQSOs is
dominated by quasars with 0, 1 or 2 \civ doublets. For example, in
the largest $\beta$-interval employed (0.00-0.05), i.e. worst-case,
there are 7550 outflow \civ doublets in total, with 4431, 992 and 241
quasars possessing 1, 2 and $>$2 \civ doublets respectively. A
significant fraction of the 241 quasars ($\simeq$150) are predicted to
result from the presence of three intervening doublets or two
intervening doublets and one outflow doublet, for which no
correction has [deliberately] been applied. In summary, the corrected
fractions of quasars exhibiting line-locked \civ absorbers are likely
to be systematically too high but by an amount that is less than than
the quoted errors (derived using counting statistics alone).

The statistics presented in Tables~\ref{tab:qstat_nbal} and
\ref{tab:qstat_bal} represent key results of the investigation:
approximately two thirds of quasars with multiple \civ absorption
systems in outflows extending to $\sim$12\,000\kms~possess line-locked
\civ absorbers, visible as \civ absorber triplets. The line-locked
absorbers are present in both non-BALQSO and BALQSO samples with
comparable frequencies.

\section{Properties of line-locked systems}\label{sec:ll_prop}

The low S/N of individual absorbers precludes their use to
investigate any systematic change in properties as a function of
$\beta$, or other parameters of interest. Information can be extracted
using composite spectra of many absorbers but first it is necessary to
isolate the absorption-line spectrum from the underlying quasar
SEDs, which possess an extended range of properties. An effective
solution is to construct a high-S/N composite quasar spectrum for each
individual quasar that possesses an absorption system. The composite
quasar spectrum can then be divided into the individual quasar
spectrum, removing the `quasar signal' and leaving the absorption-line
spectrum alone. Groups of such absorption-line only spectra can then
be combined to produce a high-S/N composite absorption-spectrum.

\begin{figure*}
\includegraphics[width=188mm]{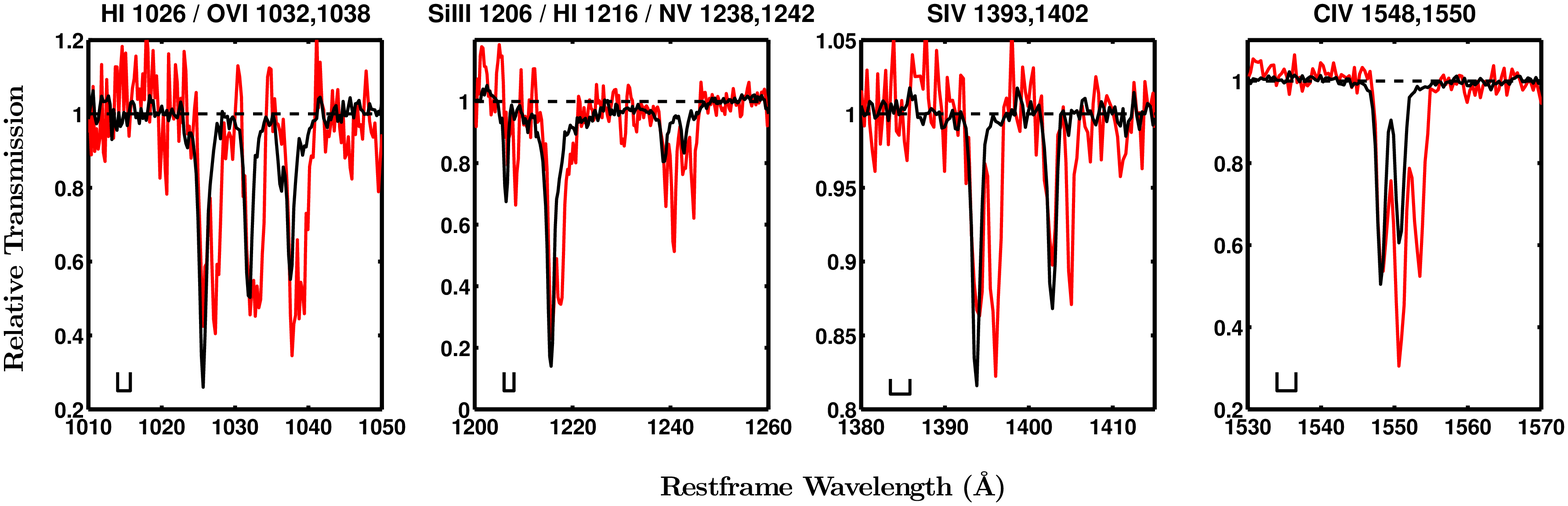}
 \caption{Arithmetic-mean composite absorption-line spectra of \civ
triplet absorbers (red) with the composite absorption-line spectra of
\civ doublets in the non-BALQSO sample overplotted (black). the
sample of 36 triplet absorbers and 448 doublet absorbers are from
quasars with 2.8$\le$$z$$\le$3.0, where the high-ionization
\ovi\ll1031.9,1037.6 doublet (ionization potential 138\,eV) is visible
in the SDSS spectra. The S/N of the triplet-composite is not high but
the presence of `double' \ovi, \hins, \nvns and \siiv absorbers,
separated by the 500\kms~\civ interval (indicated by the lines in the
bottom lefthand corner of each panel), confirms that the absorbers
consist of gas, at velocities coincident to $\le$70\kms, with physical
conditions capable of producing absorption due to species with a very
extended range of ionization potentials.
}
 \label{f:comp_ovi}
\end{figure*}

\begin{table*}
\caption{Absorption line ratios as a function of $\beta$ for the
  non-BALQSO sample, derived 
from the composite spectra described in Section~\ref{sec:ll_prop}.}
\label{tab:absrat}
\centering
%\begin{scriptsize}
\begin{tabular}{ccccc}\\ 
\hline
\multirow{1}{*}{} & \multicolumn{2}{c}{Single Doublets} &
\multicolumn{2}{c}{Triplets} \\

\multicolumn{1}{c}{$\beta$} & 
\multicolumn{1}{c}{N\,{\sc v}$_{1239}$:C\,{\sc iv}$_{1548}$} &
\multicolumn{1}{c}{Mg\,{\sc ii}$_{2796}$:C\,{\sc iv}$_{1548}$} &
\multicolumn{1}{c}{N\,{\sc v}$_{1239}$:C\,{\sc iv}$_{1548}$} &
\multicolumn{1}{c}{Mg\,{\sc ii}$_{2796}$:C\,{\sc iv}$_{1548}$}  \\
\hline
0.00-0.01   &  0.59$\pm$0.05 & 0.11$\pm$0.05 & 0.61$\pm$0.05 & 0.13$\pm$0.05 \\
0.01-0.02   &  0.57$\pm$0.05 & 0.22$\pm$0.05 & 0.54$\pm$0.05 & 0.25$\pm$0.05 \\
0.02-0.03   &  0.31$\pm$0.05 & 0.42$\pm$0.05 & $<$0.39+0.15 & 0.39$\pm$0.05 \\
0.03-0.04   &  0.20$\pm$0.05 & 0.73$\pm$0.05 & $-$ & $-$ \\
0.04-0.05   &  $<$0.01+0.02 & 0.57$\pm$0.05 & $-$ & $-$ \\
0.02-0.05   &  0.10$\pm$0.05 & 0.63$\pm$0.05 & 0.26$\pm$0.10 & 0.59$\pm$0.05 \\
Intervening &  $<$0.01+0.02 & 0.62$\pm$0.05 & $-$ & $-$ \\

\end{tabular}
\begin{minipage}{165mm}
\end{minipage}
%\end{scriptsize}
\end{table*}

\begin{figure*}
\includegraphics[width=188mm]{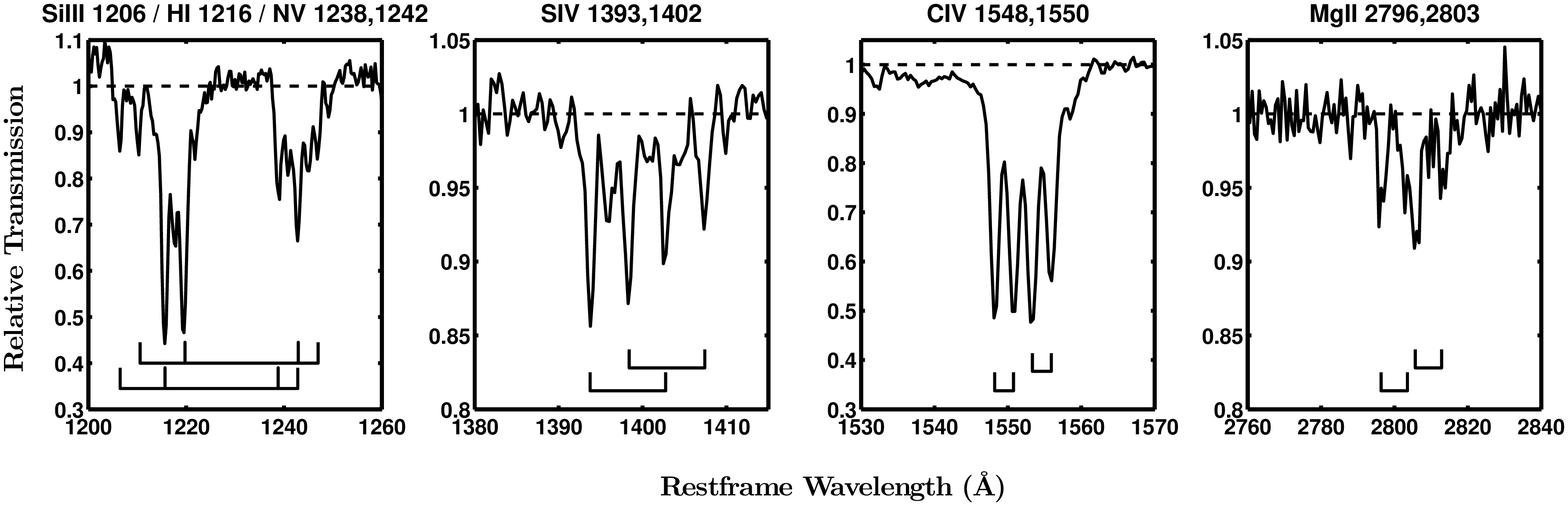}
\caption{Arithmetic-mean composite absorption-line spectra of the 159
\civ quadruple absorbers (Table~\ref{tab:lstat_nbal}) in the
non-BALQSO sample. The locations of absorption features assuming that
the quadruples are made up of two \civ doublets, with no line-locking,
are indicated by the bars at the bottom of each panel. The strong
absorption features present for each species are consistent with the
occurrence of two \civ doublets without any line-locking, in agreement
with the statistics in Table~\ref{tab:lstat_nbal}. A fraction of the
individual \civ absorbers are themselves almost certainly line-locked
(producing a triplet signature in the other absorber species) as can
be seen particularly in the \siiv panel and also with the weak fifth
absorption feature redward of the four strong absorption dips in the
\civ panel.  }
 \label{f:quad_abs}
\end{figure*}

Quasars from the non-BALQSO sample (Section~\ref{sec:quasarsample}),
without detected \civ absorption, were employed to provide a catalogue
of `control quasars', from which high-S/N composite quasars were
generated. For each quasar with a detected \civ absorber (absorber
quasar), the 100 quasars in the control sample with the closest
absolute magnitudes, $M_i$, within $\Delta z\pm$0.1 of the absorber
quasar, are identified. Selection of quasars with very similar $M_i$
minimises any systematic differences in emission line to continuum
ratios \citep{1977ApJ...214..679B}, while the narrow redshift interval
maximises the rest-wavelength range in common with the absorber
quasar. The quasars are normalised to possess the same integrated
flux, using the rest-wavelength interval 1650-2250\,\AA. The
normalised spectra are then combined, employing a simple arithmetic
mean at each rest-wavelength, to produce a composite spectrum. The
composite is normalised to the integrated flux of the absorber quasar
and divided into the absorber quasar. The result is essentially a
continuum-divided spectrum with the quasar signal removed and the
absorption systems remaining. Small, large-scale, systematic
differences in the slopes of the SEDs of the absorber and control
quasars, consistent with the presence of dust at the level of
$E(B-V)$$\simeq$0.01\, mag, associated with the absorbers or the
quasars that possess absorbers, are still evident in the absorber
composites. The implications for the dust content of the absorbers are
discussed below (Section~\ref{s:res_qseds}) but, to compare the
absorber properties of the composites, any slope differences have been
removed. Specifically, the composites have been normalised to unity
using a large-scale median-filter derived continuum (filter-scale of
251 pixels, $\simeq$100\,\AA).

\subsection{Properties of the outflowing absorbers}

Fig.~\ref{f:comp_ds} shows the results of the composite-absorber
spectrum-generation scheme applied to the sample of quasars that
possess only a single \civ doublet in the interval
0.0$\le$$\beta$$\le$0.1. Composite spectra are constructed
as a function of $\beta$. The number of individual doublets
is large, resulting in composites with a high S/N.

The single-doublet composite spectra are overplotted on the
\civns-triplet composite spectra in Fig.~\ref{f:comp_tds}. The figure
demonstrates the reality of the line-locked systems unambiguously,
through the presence of two absorber signatures (for multiple absorber
species) separated by the 500\kms~\civns-doublet interval.  Transitions
resulting from hydrogen, magnesium, nitrogen and silicon are all
present. The relative strengths of the twin absorption systems (see
for example the \siiv\ll 1393,1402 panels) is consistent with the
relatively low contamination of the observed \civ triplet sample,
predicted from the statistics presented in Table~\ref{tab:lstat_nbal}.

Figure~\ref{f:comp_tdsbals} shows the equivalent plot for the BALQSO
sample. The large-scale depressions in the `continua' evident in the
figure are due to the presence of the extended BAL-troughs in the
quasar spectra. Again, even with the poorer S/N, the signature of the
twin absorption systems making up the \civns-triplets are clearly
visible in the \hi\,$\lambda$ 1216, \nv\,\ll1238,1242 and \siiv\,\ll1393,1402
transitions. The S/N for the \mgii\,\ll2796,2803 absorption is
particularly poor but even here, the presence of twin absorber systems
can be made out at small $\beta$.

Figs.~\ref{f:comp_tds} and \ref{f:comp_tdsbals} provide a visual
representation of some of the main results of the paper: i) a large
fraction of intrinsically narrow ($\sigma$$\la$200\kms) absorption
systems, present in outflows extending to at least $\ga$20\,000\kms,
are line-locked, with velocity differences equivalent to the
separation of the \civ\,\ll1548.19,1550.77 absorption doublet, ii) the gas
associated with the narrow absorption systems produces strong
absorption due to species with a wide range of ionization potential
(15.0\,eV (\mgiins) to 97.9 (\nvns) in Figs.~\ref{f:comp_tds} and
\ref{f:comp_tdsbals}), iii) line-locked absorption systems are present
in both non-BALQSO and BALQSO samples and the dependence of the
absorber properties as a function of outflow velocity, $\beta$, is
similar.

The presence of species over an even greater ionization potential
range is illustrated by the composite absorption-line spectra of \civ
triplet absorbers and \civ doublets in the non-BALQSO sample where the
absorber redshifts allow the rest-frame spectra to extend down to
$\sim$1000\,\AA \ (Fig.~\ref{f:comp_ovi}). The sample of 36 triplet
absorbers and 448 doublet absorbers are from quasars with
2.8$\le$$z$$\le$3.0, allowing the high-ionization \ovi\ll1031.9,1037.6
doublet (ionization potential 138\,eV) to be included in the SDSS
spectra. The S/N of the triplet-composite is not high but the presence
of `double' \ovi, \hins, \nv and \siiv absorbers, separated by the
500\kms~\civns-doublet interval confirms that the absorbers consist of
gas, at velocities coincident to $<$69\kms~(i.e. one pixel in the SDSS
spectra), with physical conditions capable of producing absorption due
to species with a very extended range of ionization potentials.

The systematic trends in the properties of both the \civ doublet and
triplet absorbers, notably the strength of the high- and
low-ionization species as a function of $\beta$ seen in
Fig.~\ref{f:comp_tds}, are quantified in Table~\ref{tab:absrat}. The
ratio of the EW of \nvns$_{1239}$ to \civns$_{1548}$ decreases with
increasing $\beta$, while that of \mgiins$_{2796}$ to \civns$_{1548}$
increases. For the single \civ doublets, it is necessary to perform a
correction to the observed EW-ratios to allow for the presence of a
significant number of intervening absorbers that are unrelated to
outflows.

The EW-ratios for a composite spectrum constructed from a sample of
2200 \civ absorbers with 0.15$<$$\beta$$<$0.25 were used to provide
the values for intervening absorption systems. The EW-ratios for the
single \civns-doublet absorber-composite in the interval
0.05$<$$\beta$$\le$0.10 are indistinguishable from the intervening
absorber-composite, as expected, given intervening absorbers dominate
the sample of single doublets at $\beta$$>$0.05.  Estimates of the
\mgiins:\civ and \nvns:\civ EW-ratios in the single \civ outflowing
absorbers as a function of $\beta$ are given in
Table~\ref{tab:absrat}. The calculation assumes that the fraction of
intervening single doublet absorbers in each $\beta$ interval is given
by the predicted numbers from Section~\ref{ssec:ll_stats_orig}, after
the statistical removal of the number of quasars with multiple
intervening absorbers and intervening plus outflow absorbers that
contribute to the multiple-doublet sample (Table~\ref{tab:qstat_nbal}).
For $\beta$$>$0.03 the S/N of the triplet-absorber composites is very
low and results for the larger interval 0.02$<$$\beta$$\le$0.05 are
also given.

The systematic changes in the \mgiins:\civ and \nvns:\civ EW-ratios as
a function of $\beta$ for the single-doublet and triplet absorbers in
the outflows are consistent within the errors. The uncertainties in
the EW-ratios for all but the very weak \nvns-absorption at large
$\beta$ are dominated by uncertainties in the exact placement of the
`continua', necessary for the measurements. The poor S/N in the
absorber-composites for the BALQSO-sample precludes a similar
quantitative comparison but inspection of Fig.~\ref{f:comp_tdsbals}
suggests that the same trends as seen in the non-BALQSO sample are
present\footnote{The number of quasars with multiple \civ absorbers in the
$\Delta\beta$=0.01 intervals (for both the non-BALQSO and BALQSO
samples), even before correction for the fraction of intervening
absorbers, is too small to provide significant constraints on the
EW-ratio trends as a function of $\beta$, although the low-S/N
composites are consistent with the results for the single absorbers
and the triplets.}.

\subsection{The SEDs of the host quasars}\label{s:res_qseds}

\subsubsection{The dust content of the absorbers}\label{s:res_qseds_dust}

Interpretation of any differences in the overall shape of the
ultra-violet SEDs of the quasars with and without outflowing \civ
absorbers is not straightforward, given the presence of populations of
both associated and intervening absorbers in the same quasars. To test
for the presence of dust associated with the outflow absorbers,
composite host-quasar spectra were constructed for samples of triplet
and single doublet absorbers in the non-BALQSO sample that lie in the
interval 0.0067$<$$\beta$$<$0.0233 (i.e. 2000-7000\kms, where few
`associated'-absorbers are present). A large sample of quasars, with the
same redshift and absolute magnitude distribution, from the sample of
quasars with no detected \civ absorbers was used to provide the `control'
SED. Dividing each absorber-composite by the control-composite and
fitting the resulting ratio (over the wavelength interval
1600-3300\,\AA) with an SMC-like extinction curve gives
$E(B-V)$=0.008$\pm$0.002 and 0.016$\pm$0.003 for the doublet- and
triplet-composites respectively. The uncertainties are dominated by
the intrinsic dispersion among the constituent quasar
SEDs. The measurements are consistent with absorbers in the specified
outflow velocity-range possessing a small dust content and the triplet
sample consisting of pairs of such doublets.

The result strongly suggests that dust associated with the outflow
absorbers is responsible for the small differences in the slope of the
quasar ultraviolet SEDs but the interpretation is not unambiguous.

\subsubsection{Differences in the intrinsic SEDs of the host quasars}

\citet{2013MNRAS.432.1525B} investigated BALQSO ultraviolet SEDs and
correlations with the high-ionization absorption properties, including
outflow-velocity, $\beta$. \citet{2013MNRAS.432.1525B} draw particular
attention to the anti-correlation between the strength of the
\heii\,$\lambda $ 1640 recombination emission feature in the quasar spectra and
the increasing $\beta$ to which the BAL-troughs extend. The strength
of \heii is a measure of the far-ultraviolet quasar SED (capable of
ionizing helium). A strong far-ultraviolet SED in the quasars results
in highly ionized outflowing material and line-driven acceleration
(e.g. via \civns) is thus reduced. As a consequence, material does not
reach as extreme outflow velocities as in outflows associated with
quasars possessing weaker far-ultraviolet SEDs.

Quantifying the EW of the relatively weak \heii\,$\lambda$1640
emission line is difficult. As described in their section~2,
\citet{2013MNRAS.432.1525B} choose to define a continuum by
extrapolating from a point redward (at 1720\,\AA) of \heiins,
producing EW-values of up to 10\,\AA. Variations in the overall shape
of the quasar SED also result in some dispersion in the inferred \heii
EWs.

The continuum-definition procedure is very different from that often
employed, which results in \heiins-EW measurements of
$\simeq$0.5\,\AA, a factor of ten smaller
\citep[e.g.][]{2001AJ....122..549V}. We have adopted the more traditional
continuum-definition scheme, using a linear interpolation between the
continuum levels at 1620\AA \ and 1650\,\AA. The \heiins-EW in
composite spectra made from samples of $\simeq$1000 quasars from the
no-absorber non-BALQSO control sample is 0.62$\pm0.01$\,\AA. Dividing
the absorber sample at $\beta$=0.02, to define `close' ($<$6000\kms)
and `far' ($>$6000\kms) absorption, produces sub-samples of single
doublet (sd), multiple doublet (md) and triplet (t) absorbers that
produce composite quasar spectra with adequate S/N to measure the
relative \heiins-emission line strength to high accuracy.

The EW measurements for the close/far composites in the sd, md and t
samples are 0.70/0.57, 0.60/0.44, 0.71/0.44\,\AA \ respectively. The
continuum-definition procedure is well-defined and identical for each
composite spectrum. The ratios of the EWs in the close to far samples
are determined to an accuracy of better than 10 per cent and the three
close/far ratios, of 1.23, 1.36 and 1.61, in the same sense (stronger
\heiins-EW for absorbers with relatively low outflow velocities) as
that determined by \citet{2013MNRAS.432.1525B} is potentially
significant. The caveat is that the dominant uncertainties on the
ratios for the md and t samples is the small number of quasars in the
`far'-sub-samples and a more robust measurement is required to establish
the systematic EW-difference definitively.

\subsubsection{Differences in the global properties of the host quasars}

\citet{2011ApJS..194...45S} provide a compilation of quasar properties
for the SDSS DR7 quasars, including radio detections and estimates of
the quasar luminosity relative to the Eddington luminosity. The
non-BALQSO quasar sample can be divided into sub-populations with i)
no \civ absorbers, ii) outflow absorbers, iii) line-locked outflow
absorbers. While there are some weak trends that derive from the
detectability of absorbers as a function of the S/N in the SDSS
spectra, there are no significant differences among the
sub-populations in a) radio-detection fraction, b) quasar luminosity
or c) luminosity relative to the Eddington luminosity.

Given that a substantial fraction of all quasars show evidence for
outflows \citep[e.g.][]{2008ApJ...672..102G} and that such a large
fraction of quasars with outflows show evidence for the presence of 
line-locked absorbers, the lack of significant differences in the
distribution of such properties is perhaps not surprising.

\subsection{Constraints on other line-locked absorber 
species}\label{s:res_qseds_noll}

The focus of the investigation has been the presence of absorbers
line-locked at the 500\kms~separation of the \civ\,\ll1548,1550
doublet. The absorber detection procedure described in
Section~\ref{sec:discabs} requires absorbers to exhibit significant
\civ absorption but the `triplet'-identification scheme is sensitive
to any excess of \civ doublets present at line-locked velocity
separations of other absorption species, e.g. \nv\.\ll1238,1242
(962\kms) or \siiv\,\ll1393,1402 (1932\kms). Inspection of
Fig.~\ref{f:pixlag} and Fig.~\ref{f:pixlag_2} demonstrates that no
additional excess of absorber pairs is detectable at any separation
[that can't be attributed to pairs from different species in the same
absorber system, such as the \alii and \feii features seen in
Fig.~\ref{f:pixlag}].

In fact, the \civ doublet catalogue does not provide much sensitivity
to the detection of \civ doublets line-locked at velocity separations
$\ga$1100\kms~because there are so few quasars that possess pairs of
\civ doublets that are separated by such velocities within individual
outflows. Specifically, within the $\beta$-range 0.0$<$$\beta$$<$0.04
there are only 1543 \civ doublet outflow-pairs with velocity
separations $>$1100\kms. Almost exactly two-thirds of the pairs are
due to intervening-outflow absorber and intervening-intervening
absorber coincidences, leaving just $\simeq$500 \civ doublet
outflow-pairs. As a result, to be detectable, a line-lock at a
velocity $>$1100\kms~in the non-BALQSO sample would have to include
$\simeq$50\,per cent of the \civ doublet pairs at a given velocity. No
evidence for any such line-locks is seen.
 
The number of \civ doublets present at separations where line-locking
due to \nv\,\ll1238,1242 at 962\kms occurs is larger. The velocity
separation is only half an SDSS-spectrum pixel different from the
velocity separation of any trios of line-locked \civ doublets. The
confusion with pairs of \civ doublets separated by $\simeq$500\kms~and
a small fraction of genuine quadruples due to three line-locked \civ
doublets (Fig.~\ref{f:quad_abs}) is such that, again, the constraint
on the fraction of line-locked \nv doublets is weak.

Information on the presence of other line-locked \civ doublet pairs
will improve when larger samples of absorbers are analysed (e.g. the
SDSS III quasar samples) but from the SDSS DR7 spectra the constraints
on the prevalence of line-locking at $>$1100\kms~are weak.

\section{Discussion}\label{sec:discuss}

The presence of line-locked narrow \civ absorbers in the majority of
highly ionized outflows associated with luminous quasars will likely
provide powerful constraints on the acceleration mechanism responsible
for the outflows and the physical conditions present in the flows. A
detailed investigation of models is beyond the scope of this paper but
some initial calculations, intended to illustrate several concepts, are
presented below. The intention is not to attempt to model any
particular outflow clouds in detail, a task that will be addressed 
in subsequent papers by exploring an extensive parameter space.

The radiative acceleration is given by \citep{1974MNRAS.169..279C}
\begin{equation}
a_{rad} = \int_0^{\infty} \frac{4\pi F_{\nu}}{h\nu} \frac{\kappa_{\nu}}{\rho} \frac{1}{c}\, d\nu
\label{eqn:radaccel}
\end{equation}
where $\kappa_{\nu}$ is the gas opacity per nucleon, $\rho$ is the
mean density per nucleon, and the integral is over the incident
radiation field $F_{\nu}$.  The gas opacity $\kappa_{\nu}$ is related
to the optical depth, $\tau_{\nu}$, which we use throughout
because it is directly observed, by
\begin{equation}
\tau_{\nu} = \int  \, n \kappa_{\nu}  \, dr \! 
\end{equation}
where $n$ is the nucleon density and $r$ is the cloud depth along the
line of sight.  In practice, the clouds considered below are
fairly homogeneous, so $\tau_{\nu}$ and $\kappa_{\nu}$ are nearly
proportional to one another.

Two dependencies are revealed by equation \ref{eqn:radaccel}.
First, the metallicity $Z$ enters through the ratio $ \kappa_{\nu} / {\rho}$ because heavy elements provide most of
the opacity but little of the mass - this ratio is proportional to $Z$.
We assume conventional metallicities and do not explore consequences of changing $Z$ in this paper.
Second, the SED of the AGN enters through $F_{\nu}$.  
We assume that the ratio $4\pi F_{\nu} / h\nu$ is constant as a first approximation although
other SEDs are considered in Section~\ref{sec:agn_shape}.

\subsection{Baseline \civ clouds}\label{sec:clouds}

Version 13.03 of {\sc Cloudy}, most recently described by
\citet{2013RMxAA..49..137F}, is employed to simulate conditions within
clouds.  The results given in table~3 of \citet{2011MNRAS.410.1957H}
are used as a guide to set up a simple baseline photoionization model.
It is assumed that the cloud is irradiated by the standard active
galactic nuclei (AGN) SED built into {\sc Cloudy}.

Gas found near the centres of AGN, or massive galaxies generally, has
metallicities significantly above solar
\citep{1999ARA&A..37..487H}. Therefore, adopting a conservative
approach, metallicity, $Z$, is set at three times higher than solar.
In practice, the default solar abundance for all elements heavier than
helium is increased by a factor of three, with nitrogen increased by a
further factor of three to account for secondary nitrogen production.

A representative column density of $N$(\civns)=10$^{14}$\cm2 and a
microturbulent velocity of 50\kms are assumed, along with
a hydrogen density of $10^5$ cm$^{-3}$.  
This turbulence is supersonic with a Mach number of $\simeq 4$, 
meaning that turbulent pressure will dominate over gas pressure.
These parameters are
typical of the structures studied in \citet{2011MNRAS.410.1957H} and
produce an optical depth in the \civ doublet of a few, close to the
observed value and optimised to produce the maximum acceleration due
to \civns.  The cloud is roughly isothermal with a temperature of
$T$$\sim$20\,000\,K, a typical value.

\subsection{Hydrogen column density and cloud reddening}

Photoionization simulations of the absorbers cannot determine the
density of the cloud, only its ionization parameter $U$ (i.e. the
number of ionizing photons per atom).  The results above
(Section~\ref{s:res_qseds_dust}) suggest that each cloud has an
internal reddening of $E(B-V)\approx 0.008$ mag.  The grain optical
properties, the ratio of total to selective extinction $R$, and the
dust to gas ratio, are unknown. \citet{1991ApJ...379..245S} find
\begin{equation}
\frac{N({\rm H})}{E(B-V)} = 5 \times 10^{21} \ {\rm cm}^{-2} \ \rm{mag}^{-1} .
\end{equation}
to be typical within our Galaxy.  If the dust to gas ratio scales with
the metallicity $Z$, then the hydrogen column density corresponding to
the deduced reddening is
\begin{equation}
N({\rm H}) = 4 \times 10^{19} \ {\rm cm}^{-2} \ Z^{-1} 
\end{equation}

\begin{figure}
\includegraphics[width=80mm]{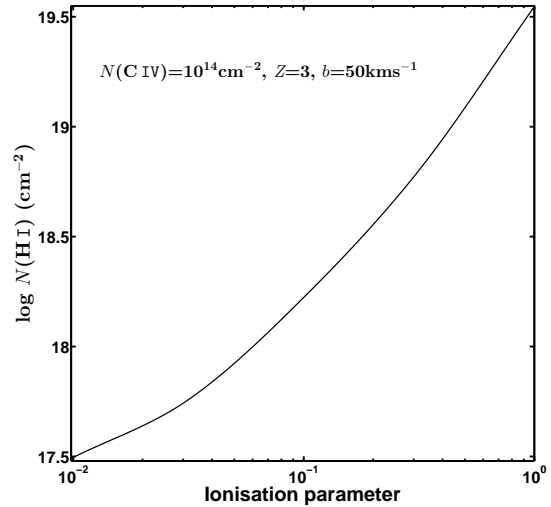}
\caption{The total hydrogen column density $N$(H) is shown as a
function of the dimensionless ionization parameter $U$.  Enhanced
abundances, a typical AGN SED, and a microturbulent velocity of 50\kms
were assumed.  The measured reddening of 0.008 mag/cloud suggests log
$U$ $\sim -0.3$}
\label{fig:varyU}
\end{figure}

A series of models are computed, where the ionization parameter, $U$,
varies and the cloud thickness is adjusted to keep
$N$(\civns)=10$^{14}$\cm2.  Figure~\ref{fig:varyU} shows the total
hydrogen column density as a function of $U$.

The deduced reddening and assumed grain properties correspond to log
$U \sim -0.3$ for $Z$=3.  We adopt this abundance and value of $U$ in
the following work.

\subsection{Line optical depths for the fiducial cloud}

\begin{figure}
\includegraphics[width=80mm]{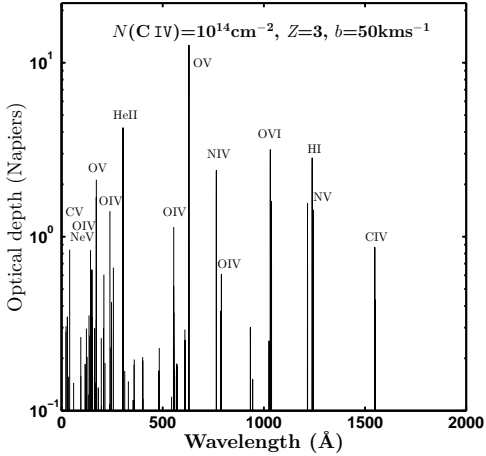}
\caption{The line optical depths of the fiducial model are shown.  The
gas is fairly highly ionized, with most elements having four to six
electrons removed, and the resulting permitted lines are mainly in the
extreme ultraviolet.  The very shortest wavelength lines are
inner-shell transitions.  The \civ doublet components are the
longest-wavelength lines. The lines indicated at the shortest,
very crowded, wavelengths are \cv$\lambda$40, \nev$\lambda$143, \oiv$\lambda$150 and \ov$\lambda$172. }
\label{fig:OpticalDepths}
\end{figure}

{\sc Cloudy} includes a high-resolution fine opacity grid that is used for
multi-grid calculations of line acceleration and overlap
\citep{2013RMxAA..49..137F}.  The predicted line optical depths in
this fine grid are shown in Fig.~\ref{fig:OpticalDepths}.

Individual lines are well resolved on this fine grid, which has $R
\equiv \lambda / {\delta \lambda} \sim 10^4$. In a highly ionized gas
with enhanced abundances, the largest opacities are due to resonance
lines of Li-like, He-like and H-like ions of O, C and N.  The \heii
and \hi resonance lines are also significant.  There are an especially
large number of lines at short wavelengths, $\leq 300$\AA.  The lines
grow sparser at longer wavelengths and the \civ doublet is the
longest-wavelength feature.  The details of the model do not affect
these qualitative features.

\subsection{The line autocorrelation signature}

\begin{figure}
\begin{center}
\includegraphics[width=80mm]{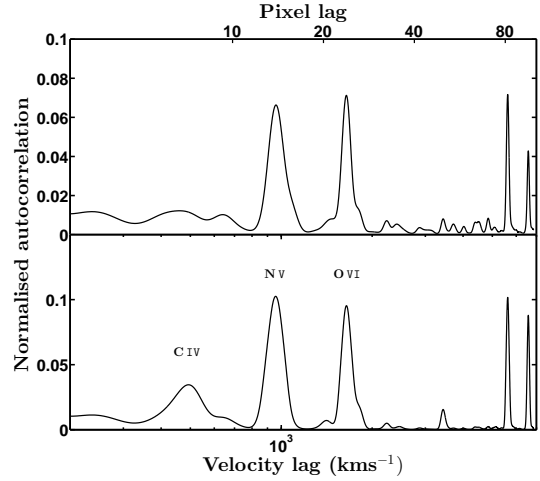}
\end{center}
\caption{The line optical depth autocorrelation function is shown for
the fiducial model.  The top panel shows results for the lines shown
in Fig.~\ref{fig:OpticalDepths} with a maximum optical depth of
$\tau$=2. The lower panel shows the autocorrelation for the same lines
but weighted by the $\nu^{-0.7}$ factor, approximating the form of the
incident AGN SED (see Section~\ref{sec:agn_shape}); the \civ doublet
becomes the strongest feature below $\simeq$1000\kms. The positions of
the strong \civ, \nv, and \ovi doublet pairs are indicated in the
lower panel. The ACFs are normalised such that they have a value of
unity at zero-velocity and the features of interest are the relative
heights of the peaks at velocities greater than 200\kms (the minimum velocity shown on the x-axis above). The axis at
the top shows the pixel-lag, to allow direct comparison with the
observational data shown in Figs.~\ref{f:pixlag} and \ref{f:pixlag_2}.}
\label{fig:acorr}
\end{figure}

Clouds will tend to be accelerated up to line-locked velocities if
line radiative acceleration is the dominant force acting on the
clouds, and if the mutual self-shielding of overlapping strong lines
produces a local minimum in the acceleration.  The easiest way to
search for such overlaps is by computing the line optical-depth
autocorrelation function (ACF). The optimal situation for line-driven
radiative acceleration is when the clouds possess an optical depth,
$\tau_{\nu}$, of order unity for the line species. The momentum transfer
from a line species essentially saturates once the optical depth is a
few (there are hardly any photons remaining to be absorbed) and before
calculating the ACF a ceiling of $\tau_{\nu}=2$ was applied to the line
optical depths shown in Fig.~\ref{fig:OpticalDepths}.  The results of
an autocorrelation analysis of the opacity distribution, is shown in
the top panel of Fig.~\ref{fig:acorr}.

The autocorrelation results shown extend to velocity-separations of
$\sim$7000\kms and, as discussed in Section~\ref{s:res_qseds_noll},
the observational constraints on the presence of line-locks at
separations $\ga$1000\kms are poor. That said, the form of the ACF at
lower velocities is not initially encouraging, as the 500\kms signal
due to the \civ doublet is immersed among a number of other line-lock
signatures with similar velocities, due primarily to line species in
the far ultraviolet part of the AGN SED.

For reasons related to the finite resolution of the SDSS-spectra and
the velocity-separation of the \civ doublet
(Section~\ref{sec:discabs}) identification of clouds with detectable
\civ doublets, line-locked at separations $<$500\kms is not
possible. At velocities $>$500\kms however, there is no reason why
pairs of \civ doublets would not be found with velocity separations in
the range 500-1000\kms. Such pairs would manifest themselves as an
excess number of absorber detections above the background for pixel
lags $\simeq$15-30 in the left-hand panel of
Fig.~\ref{f:pixlag_2}. Any such population present is clearly at a
much lower incidence than is the case for absorbers line-locked at the
\civ doublet separation. The well-defined narrow peak of the third
\civ absorption component evident in Fig.~\ref{f:pixlag_2} and the
essentially identical velocity-widths of the two intrinsic components
making up each \civ line-locked triplet (Fig.~\ref{f:comp_tds}) also
provide evidence against the presence of a significant population of
additional line-locked \civ absorbers with velocity separations
500-1000\kms. In principle, other line-pairings, with velocity
separations of $\simeq$500\kms, could be contributing to the peak at
500\kms that we have ascribed to \civns. Removing the \civns-doublet
from the absorption-line opacity table, however, demonstrates that
\civ dominates the contribution to the 500\kms peak completely.  A
focus on factors that result in an enhanced \civ doublet signal,
consistent with the observations, therefore appears worthwhile.

\subsubsection{Changing the AGN SED Shape}\label{sec:agn_shape}

The analysis above assumes a uniform weighting of the lines, ignoring
the shape of the incident AGN SED responsible for accelerating the gas.

The SED used to compute Figs.~\ref{fig:OpticalDepths} and
\ref{fig:acorr} (top panel) is a generic AGN SED, deduced in part from
observations of intrinsic emission lines.  The SED that drives the
absorbing clouds may not be the same.  The central regions of an AGN
are believed to have a cylindrical geometry with the emitting clouds
along the equator.  It is likely that we are viewing the AGN from
regions near the axis of the cylinder since we have a relatively
unobstructed view of the SED.

To examine the effects of various SEDs upon the auto correlation
function the SED may be parametrized as a single power-law.  The
approach is only valid in the broadest sense, but has the advantage of
simplicity and serves to show the dependencies of the ACF on the
illuminating SED.

Assume therefore, that the AGN SED can be fitted, over the wavelength
range shown in Fig.~\ref{fig:OpticalDepths}, with a power-law as a
function of frequency, $\nu$, $f_{\nu} \propto \nu^{\alpha}$.  The
photon-number flux is then $n_{\nu} \propto \nu^{(\alpha-1)}$.  Assume
further that lines are microturbulently broadened to a velocity width
$du$ (50\kms is adopted in the baseline model).  The line width
in frequency units is $\delta \nu = du \nu /c$, where $c$ is the
velocity of light, and the number of photons within an absorption line
is $n_{line} = n_{\nu} \delta \nu \propto \nu^{\alpha}$.  Each photon
has momentum $h \nu/c$, so the photon momentum transferred to the
clouds is $\propto n_{line} h \nu/c \propto \nu^{\alpha + 1}$.

The main effect of varying the slope of the SED is to change the
weighting given to lines at different wavelengths.  Steep slopes give
greater weight to lines at longer wavelength.  We adopt a power-law
index of $\alpha$=-1.7, typical for the SDSS-quasars in the sample
\citep[e.g.][]{2011AJ....142..130K}. The optical depths shown in
Fig.~\ref{fig:OpticalDepths} were rescaled by a factor
$\propto\lambda^{0.7}$ and normalised to give a maximum optical-depth
of $\tau=2$ for the \civ 1548\,\AA \ line. The \civ doublet optical
depth for the sample has to be of this order for the lines to be
detected.

The resulting ACF is shown in the lower panel of
Fig.~\ref{fig:acorr}. The steeper AGN SED results in the
long-wavelength line pairs, due to \ovins, \nv and \civns, becoming
stronger. In particular, the differences at velocities $\la$1000~\kms,
compared to the unweighted case, are significant and the peak due to
the \civ doublet line-lock is much more pronounced and more consistent
with the observations.

Future work should investigate how the ACF depends on the SED.
Sophisticated AGN SEDs including the effects of a range in $L/L(Edd)$
now exist \citep{2012MNRAS.425..907J}.  These could be combined with
the ACF approach taken in this paper to investigate how the properties
of the outflow depend on other AGN parameters.

\subsection{Radiative acceleration and cloud stability}

Line radiative acceleration must be important if line locking is to occur.
Our {\sc Cloudy} calculations include a self-consistent treatment of
acceleration by both lines and continua \citep{2013RMxAA..49..137F}.
Line self-shielding and overlap with other lines is fully treated
using the fine opacity grid shown in Fig.~\ref{fig:OpticalDepths}.

The database CLOUDY employs for resonance lines (transitions arising
from the ground state) should be fairly complete and such transitions
dominate the line-dependent acceleration. The situation contrasts with
that applying to stellar winds where \citet{1975ApJ...195..157C}
stressed that millions of lines arising between excited states can
contribute to the acceleration in a stellar atmosphere.  This is not
the case in the low density AGN absorber material considered here,
where most atoms will be in the ground state.  In an O-star wind,
collisional processes resulting from the high densities create a
substantial population of atoms in excited states, so subordinate lines (transitions between two excited states) are important.  There are
\emph{many} more permitted subordinate lines, which are included in
the CLOUDY calculations, but their influence is found to be
unimportant.  Fig.~\ref{fig:OpticalDepths} shows all lines with
significant optical depth.

\begin{figure}
\begin{center}
\includegraphics[width=80mm]{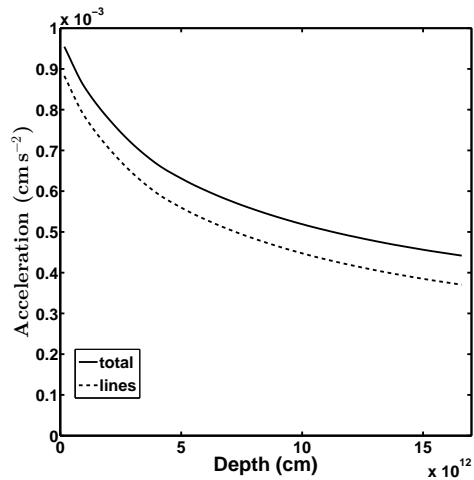}
\end{center}
\caption{The radiative acceleration as a function of cloud depth for
the fiducial model.  The total- and line-only accelerations are the
upper and lower curves respectively.}
\label{fig:AccellDepth}
\end{figure}

Fig.~\ref{fig:AccellDepth} shows the acceleration across the fiducial
cloud.
The upper curve shows the total radiative acceleration while the lower
curve includes only lines.  The radiative acceleration is
predominantly due to lines across the entire cloud.  The acceleration
is greatest at the illuminated face of the cloud, a depth of zero.
The acceleration decreases as lines become optically thick and are
self-shielded. As shown by \citet[][equation
3.48]{1974MNRAS.169..279C}, the radiative acceleration decreases when
the line optical depth becomes significant.  The outward acceleration
has fallen by about a factor of two at the shielded face of the cloud.

Although lines contribute 89 per cent of the total radiative
acceleration, the \civ doublet is not an important contributor to this
total.  The \ov$\lambda$629.7 line contributes over two thirds of the
total acceleration, with lesser contributors from the \ovi and \nv
doublets.  The \civ doublet contributes only about 1 per cent of the total.
Nearly all of the acceleration due to continuum absorption is due to
photoelectric opacity, with electron scattering being negligible.

The exploratory investigation has been successful in showing that
line-driving is capable of accelerating `clouds' with properties given
in Section~\ref{sec:clouds}. In the context of the observations, it
has also been possible to show that the line-locking signature due to
\civ exists as a local maximum in velocity space. In terms of the
spectrum of potential line-locks, the 500\kms feature
represents the lowest velocity feature present and the idea that material
in an outflow might `lock' naturally at the first significant peak in
the line autocorrelation function (Fig.~\ref{fig:acorr}; lower panel)
is attractive. What remains unclear, however, is why material in
outflows appears to maintain the 500\kms separation when the
contribution of the \civns-doublet represents such a small fraction of
the total acceleration.

Line-driven winds are famously unstable \citep{1988ApJ...335..914O}.
The decreasing acceleration means that the cloud will be compressed,
becoming a pancake \citep{1982ApJ...252...39M}.  Eventually, the the
geometry shown in Fig.~\ref{fig:AccellDepth} will become
Rayleigh-Taylor unstable and will fragment over scales similar to the
line shielding length.  This brings in rich dynamical aspects which
merit further more detailed investigation.

Fig.~\ref{fig:AccellDepth} gives hints to the origin of the line
locking phenomenon.  The properties of the cloud were taken from
previous investigations and were not `fine tuned' for this work.  Had
the column density been significantly smaller than assumed, the lines
would have been optically thin (and hence undetectable in this study)
and the cloud would also have been stable.  Had the column density
been significantly larger, the lines would have been far more
optically thick and self-shielded, with the acceleration falling
precipitously as a consequence.  Such a cloud would fragment over
scales corresponding to roughly optical depth unity in the driving
lines.  In fact, for these assumed properties, the cloud has precisely
these properties.  Could the currently visible, line-locked, small
column density clouds be the result of radiative instabilities in a
cloud which was originally considerably larger? Such a scenario might
explain why the line-locking signature for the narrow absorbers is
also present among the BALQSO sample.  Alternatively, in a model in
which BAL features are due to winds above the accretion disk, seen
only for special viewing angles, the narrow absorption systems studied
in this paper could result from the outwardly accelerated interstellar
medium of the host galaxy, visible for a much larger range of viewing
angle.

\section{Conclusions}\label{sec:conclude}

Investigation of the statistics of strong, narrow \civ doublet
absorbers present in outflows seen in a large sample of luminous
quasars (-25.0$\la$$M_i$$\la$-27.0) demonstrates that line-locked \civ
absorber doublets are an almost ubiquitous feature of outflows. From
the samples of $\simeq$31\,000 non-BAL and $\simeq$2600 BAL quasars it
is found that:

$\bullet$ Approximately two thirds of quasars with multiple \civ absorption
systems in outflows extending to $\sim$12\,000\kms possess line-locked
\civ absorbers, visible as \civ absorber triplets. 

$\bullet$ The line-locked absorbers are present in both the non-BALQSO and 
BALQSO samples with comparable frequencies, suggesting a common origin
for the acceleration of the clouds traced by the narrow \civ absorbers
in non-BAL and BAL quasars.

$\bullet$ Line-locked \civ systems are present in outflows with velocities
  reaching at least 20\,000\kms.

$\bullet$ There are no detectable differences in the absorber properties
and the dust content of single \civ doublets and line-locked \civ doublets.

$\bullet$ The gas associated with both single and line-locked \civ
narrow absorption systems produces strong absorption due to species
with a wide range of ionization potential (15\,eV (\mgiins) to 98
(\nvns) and even 138\,eV (\ovi)).

$\bullet$ Both single and line-locked \civ absorber systems show
strong systematic trends in their ionization as a function of outflow
velocity $\beta$. The ratio of the EW of \nvns$_{1239}$ to
\civns$_{1548}$ decreases with increasing $\beta$, while that of
\mgiins$_{2796}$ to \civns$_{1548}$ increases.

$\bullet$ The current sample of absorbers from the SDSS DR7 quasars does
not provide useful constraints on the prevalence of line-locks with
velocity separations $\ga$1000\kms, including those due to
\nv and \ovi at 964\kms and 1654\kms respectively.

$\bullet$ Simulations employing {\sc Cloudy} demonstrate that a rich
spectrum of line-locked signals at various velocities may be expected
due to significant opacities from resonance lines of Li-like, He-like
and H-like ions of O, C and N, along with contributions from \heii and
\hi resonance lines.

$\bullet$ Computing the line optical-depth autocorrelation function
(ACF) for the full AGN SED spectrum for parameters taken from
published models provides a quantitative way to assess the importance
of overlapping lines.

$\bullet$ Many strong contributors to the ACF are found, particular those
from \ovi, \nv and \civ doublets but, in the general case, the \civ
doublet lock is not the strongest.

$\bullet$ The strength of particular line-locks depends on the shape
of the illuminating AGN SED, which provides the photons responsible
for the acceleration at the wavelength of each absorber species. The
\civ line-lock can be made to appear significantly stronger by
adopting an AGN SED shape in the ultraviolet to near X-ray range that
is consistent with current knowledge of the AGN SED.

$\bullet$ The {\sc Cloudy} simulations confirm that line driving is
the dominant acceleration for clouds with a column of
$N$(\civns)$\simeq$10$^{14}$\cm2 and $N$(H)$\simeq$10$^{19}$\cm2.  The
outward acceleration falls by a factor of two across our fiducial
cloud, showing that the flow is Rayleigh-Taylor unstable.  Such a flow
will fragment over scales similar to that observed---the point where
the acceleration has fallen by a moderate factor.  Could the clouds we
observe be the result of such fragmentation of clouds that were
originally significantly larger?

The observational results presented open up a potentially important new
constraint for i) understanding the origin of the accelerating
mechanism for high-ionization outflows in quasars and ii) the physical
conditions pertaining within the flow. Observationally, improved
constraints on the frequency of line-locks due to other species will
be possible through the analysis of the forthcoming release (DR11) of
the SDSS III quasar catalogue. Modelling outflows is a challenging,
many-parameter, problem but the dependence of line-lock signatures
from different species on the form of the AGN SED is a particularly
promising direction for future work, on which we intend to embark. The
improved statistics from the SDSS III quasar sample will also provide
detailed determinations of the outflow cloud properties over an
extended range in $\beta$.

\section*{Acknowledgements}

We thank the anonymous referee for a constructive and thoughtful report and
Fred Hamann for comments and discussions. Bob Carswell's input, over an extended period, is much appreciated.
PCH acknowledges support from the STFC-funded Galaxy Formation and
Evolution programme at the Institute of Astronomy. JTA acknowledges
the award of an STFC Ph.D.\ studentship and an ARC Super Science
Fellowship. GJF acknowledges support by NSF (1108928, 1109061, and 1412155),
NASA (10-ATP10-0053, 10-ADAP10-0073, NNX12AH73G, and ATP13-0153), and
STScI (HST-AR-12125.01, HST-AR- 13245, GO-12560, HST-GO-12309, and
GO-13310.002-A), and is grateful to the Leverhulme Trust for support
via the award of a Visiting Professorship at Queen's University
Belfast (VP1-2012-025).

Funding for the SDSS and SDSS-II has been provided by the Alfred
P. Sloan Foundation, the Participating Institutions, the National
Science Foundation, the U.S. Department of Energy, the National
Aeronautics and Space Administration, the Japanese Monbukagakusho, the
Max Planck Society, and the Higher Education Funding Council for
England. The SDSS Web Site is http://www.sdss.org/.

The SDSS is managed by the Astrophysical Research Consortium for the
Participating Institutions. The Participating Institutions are the
American Museum of Natural History, Astrophysical Institute Potsdam,
University of Basel, University of Cambridge, Case Western Reserve
University, University of Chicago, Drexel University, Fermilab, the
Institute for Advanced Study, the Japan Participation Group, Johns
Hopkins University, the Joint Institute for Nuclear Astrophysics, the
Kavli Institute for Particle Astrophysics and Cosmology, the Korean
Scientist Group, the Chinese Academy of Sciences (LAMOST), Los Alamos
National Laboratory, the Max-Planck-Institute for Astronomy (MPIA),
the Max-Planck-Institute for Astrophysics (MPA), New Mexico State
University, Ohio State University, University of Pittsburgh,
University of Portsmouth, Princeton University, the United States
Naval Observatory, and the University of Washington.

\end{document}